\documentclass[twocolumn, floatfix, tighten]{aastex63}

\usepackage[T1]{fontenc}
\usepackage{xcolor}
\usepackage{ulem}
\usepackage{mathrsfs}

\makeatletter
\DeclareRobustCommand{\HI}{%
  \mbox{H\check@mathfonts\fontsize\sf@size\z@\selectfont I}%
}
\makeatother
\newcommand{\lya}{Ly$\alpha$}
\newcommand{\lyb}{Ly$\beta$}

\newcommand{\kms}{km~s$^{-1}$}

\newcommand{\taueff}{$\tau_{\rm eff}$}
\newcommand{\mfp}{$\lambda_{\rm mfp}$}

\newcommand{\rom}[1]{\uppercase\expandafter{\romannumeral #1\relax}}

\def\teff{\tau_{\rm eff} }

\defcitealias{becker_mean_2021}{B21}

\shorttitle{Mean free path over $5<z<6$}
\shortauthors{Zhu et al.}

\begin{document}

\title{Probing Ultra-late Reionization: Direct Measurements of the Mean Free Path over $5<z<6$}

\author[0000-0003-3307-7525]{Yongda Zhu}
\affiliation{Department of Physics \& Astronomy,
    University of California, Riverside, CA 92521, USA;
    \href{mailto:yzhu144@ucr.edu}{\emph{yzhu144@ucr.edu}}
    }

\author[0000-0003-2344-263X]{George D. Becker}
\affiliation{Department of Physics \& Astronomy,
    University of California, Riverside, CA 92521, USA;
    \href{mailto:yzhu144@ucr.edu}{\emph{yzhu144@ucr.edu}}
    }

\author[0000-0002-0421-065X]{Holly M. Christenson}
\affiliation{Department of Physics \& Astronomy,
    University of California, Riverside, CA 92521, USA;
    \href{mailto:yzhu144@ucr.edu}{\emph{yzhu144@ucr.edu}}
    }

\author[0000-0003-2344-263X]{Anson D'Aloisio}
\affiliation{Department of Physics \& Astronomy,
    University of California, Riverside, CA 92521, USA;
    \href{mailto:yzhu144@ucr.edu}{\emph{yzhu144@ucr.edu}}
    }

\author[0000-0001-8582-7012]{Sarah E. I. Bosman}
\affiliation{Max-Planck-Institut f\"{u}r Astronomie, K\"{o}nigstuhl 17, D-69117 Heidelberg, Germany}
\affiliation{Institute for Theoretical Physics, Heidelberg University, Philosophenweg 12, D-69120, Heidelberg, Germany}

\author[0000-0002-5268-2221]{Tom Bakx}
\affiliation{Department of Space, Earth and Environment, Chalmers University of Technology, Onsala Space Observatory, 439 92 Onsala, Sweden}
\affiliation{Division of Particle and Astrophysical Science, Nagoya University, Furo-cho, Chikusa-ku, Nagoya 464-8602, Japan}
\affiliation{National Astronomical Observatory of Japan, 2-21-1, Osawa, Mitaka, Tokyo 181-8588, Japan}

\author[0000-0003-3693-3091]{Valentina D'Odorico}
\affiliation{INAF-Osservatorio Astronomico di Trieste, Via Tiepolo 11, I-34143 Trieste, Italy}
\affiliation{Scuola Normale Superiore, Piazza dei Cavalieri 7, I-56126 Pisa, Italy}
\affiliation{IFPU-Institute for Fundamental Physics of the Universe, via Beirut 2, I-34151 Trieste, Italy}

\author[0000-0002-4314-021X]{Manuela Bischetti}
\affiliation{Dipartimento di Fisica, Sezione di Astronomia, Universitá di Tri- este, via Tiepolo 11, 34143 Trieste, Italy}

\author[0000-0001-9420-7384]{Christopher Cain}
\affiliation{Department of Physics \& Astronomy,
    University of California, Riverside, CA 92521, USA;
    \href{mailto:yzhu144@ucr.edu}{\emph{yzhu144@ucr.edu}}
    }

\author[0000-0003-0821-3644]{Frederick B.~Davies}
\affiliation{Max-Planck-Institut f\"{u}r Astronomie, K\"{o}nigstuhl 17, D-69117 Heidelberg, Germany}

\author[0000-0002-3324-4824]{Rebecca L. Davies}
\affiliation{Centre for Astrophysics and Supercomputing, Swinburne University of Technology, Hawthorn, Victoria 3122, Australia}
\affiliation{ARC Centre of Excellence for All Sky Astrophysics in 3 Dimensions (ASTRO 3D), Australia}

\author[0000-0003-2895-6218]{Anna-Christina Eilers}\thanks{Pappalardo Fellow}
\affiliation{MIT Kavli Institute for Astrophysics and Space Research, 77 Massachusetts Avenue, Cambridge, MA 02139, USA}

\author[0000-0003-3310-0131]{Xiaohui Fan}
\affiliation{Steward Observatory, University of Arizona, 933 N Cherry Ave, Tucson, AZ 85721, USA}

\author[0000-0002-2423-7905]{Prakash Gaikwad}
\affiliation{Max-Planck-Institut f\"{u}r Astronomie, K\"{o}nigstuhl 17, D-69117 Heidelberg, Germany}

\author[0000-0001-8443-2393]{Martin G. Haehnelt}
\affiliation{Kavli Institute for Cosmology and Institute of Astronomy, Madingley Road, Cambridge, CB3 0HA, UK}

\author[0000-0001-5211-1958]{Laura C. Keating}
\affiliation{Institute for Astronomy, University of Edinburgh, Blackford Hill, Edinburgh, EH9 3HJ, UK}

\author[0000-0001-5829-4716]{Girish Kulkarni}
\affiliation{Tata Institute of Fundamental Research, Homi Bhabha Road, Mumbai 400005, India}

\author[0000-0001-9372-4611]{Samuel Lai}
\affiliation{Research School of Astronomy and Astrophysics, Australian National University, Canberra, ACT 2611, Australia}

\author[0000-0002-5237-9433]{Hai-Xia Ma}
\affiliation{Division of Particle and Astrophysical Science, Nagoya University, Furo-cho, Chikusa-ku, Nagoya 464-8602, Japan}

\author[0000-0003-3374-1772]{Andrei Mesinger}
\affiliation{Scuola Normale Superiore, Piazza dei Cavalieri 7, I-56126 Pisa, Italy}

\author[0000-0002-4314-1810]{Yuxiang Qin}
\affiliation{School of Physics, University of Melbourne, Parkville, VIC 3010, Australia}
\affiliation{ARC Centre of Excellence for All Sky Astrophysics in 3 Dimensions (ASTRO 3D), Australia}

\author[0000-0001-5818-6838]{Sindhu Satyavolu}
\affiliation{Tata Institute of Fundamental Research, Homi Bhabha Road, Mumbai 400005, India}

\author[0000-0001-8416-7673]{Tsutomu T.\ Takeuchi}
\affiliation{Division of Particle and Astrophysical Science, Nagoya University, Furo-cho, Chikusa-ku, Nagoya 464-8602, Japan}
\affiliation{The Research Center for Statistical Machine Learning, the Institute of Statistical Mathematics, 10-3 Midori-cho, Tachikawa, Tokyo 190-8562, Japan}

\author[0000-0003-1937-0573]{Hideki Umehata}
\affiliation{Institute for Advanced Research, Nagoya University, Furocho, Chikusa, Nagoya 464-8602, Japan}
\affiliation{Department of Physics, Graduate School of Science, Nagoya University, Furocho, Chikusa, Nagoya 464-8602, Japan}

\author[0000-0001-5287-4242]{Jinyi Yang}
\affiliation{Steward Observatory, University of Arizona, 933 N Cherry Ave, Tucson, AZ 85721, USA}

\begin{abstract}
The mean free path of ionizing photons, \mfp, is a critical parameter for modeling the intergalactic medium (IGM) both during and after reionization. We present direct measurements of \mfp\ from QSO spectra over the redshift range $5<z<6$, including the first measurements at $z\simeq5.3$ and 5.6. Our sample includes data from the XQR-30 VLT large program, as well as new Keck/ESI observations of QSOs near $z \sim 5.5$, for which we also acquire new [\ion{C}{2}] 158$\mu$m redshifts with ALMA. By measuring the Lyman continuum transmission profile in stacked QSO spectra, we find $\lambda_{\rm mfp} = 9.33_{-1.80}^{+2.06}$, $5.40_{-1.40}^{+1.47}$, $3.31_{-1.34}^{+2.74}$, and $0.81_{-0.48}^{+0.73}$ pMpc at $z=5.08$, 5.31, 5.65, and 5.93, respectively. Our results demonstrate that \mfp\ increases steadily and rapidly with time over $5<z<6$. Notably, we find that \mfp\ deviates significantly from  predictions based on a fully ionized and relaxed IGM as late as $z=5.3$. By comparing our results to model predictions and indirect \mfp\ constraints based on IGM \lya\ opacity, we find that the \mfp\ evolution is consistent with scenarios wherein the IGM is still undergoing reionization and/or retains large fluctuations in the ionizing UV background well below redshift six.
\end{abstract}
\keywords{\href{http://astrothesaurus.org/uat/1383}{Reionization (1383)}, 
\href{http://astrothesaurus.org/uat/813}{Intergalactic medium (813)}, 
\href{http://astrothesaurus.org/uat/1317}{Quasar absorption line spectroscopy (1317)}, \href{http://astrothesaurus.org/uat/734}{High-redshift galaxies (734)}
}

\section{Introduction \label{sec:introduction}}

When and how reionization occurred is a fundamental question about the early universe and the first galaxies. The appearance of transmitted flux in the Lyman-$\alpha$ (\lya) forest of high-redshift QSOs \citep[e.g.,][]{fan_constraining_2006} has long been interpreted as evidence that hydrogen in the intergalactic medium (IGM) was largely reionized by $z = 6$.  In terms of the ionizing photon budget, however, an end of reionization at $z \ge 6$ is challenging to reconcile with a midpoint of $z \sim 7-8$ suggested by e.g., cosmic microwave background (CMB) observations \citep[][see also \citealp{de_belsunce_inference_2021}]{planck_collaboration_planck_2020}. In particular, star-forming galaxies at $z > 6$ would have to 
emit ionizing photons extremely efficiently 
in order to complete reionization within such a short interval.  This leaves two possibilities: the ionizing efficiency of galaxies at $z > 6$ is remarkably high, and/or reionization extends to lower redshifts.

A number of observations have now been used to constrain the timeline of reionization. These observations include the \lya\ damping wing in $z\gtrsim7$ QSO spectra \citep[e.g.,][]{banados_800-million-solar-mass_2018,davies_quantitative_2018,wang_significantly_2020,yang_poniuaena_2020,greig_igm_2021}, the decline in observed \lya\ emission from galaxies at $z>6$ \citep[e.g.,][and references therein, but see \citealp{jung_texas_2020-1,wold_lager_2021-1}]{mason_universe_2018,mason_inferences_2019,hoag_constraining_2019,hu_ly_2019}, and measurements of the thermal state of the IGM at $z>5$ \citep[e.g.,][]{boera_revealing_2019,gaikwad_consistent_2021}. These observations support a midpoint of reionization at $z\sim7-8$ and are generally consistent with an ending point at $z\sim6$, as constrained by the fraction of dark pixels in the \lya\ forest \citep[e.g.,][but see \citealp{zhu_long_2022}]{mcgreer_model-independent_2015}.

Other observations, however, suggest that reionization may have extended to significantly lower redshifts. Large-scale fluctuations in the measured \lya\ effective optical depth\footnote{Defined as $\teff=-\ln{\langle F \rangle}$, where $F$ is the continuum-normalized transmission flux.} \citep[e.g.,][]{fan_constraining_2006,becker_evidence_2015,eilers_opacity_2018,bosman_new_2018,bosman_hydrogen_2022,yang_measurements_2020}, together with long troughs extending down to or below $z\simeq5.5$ in the \lya\ forest \citep[e.g.,][]{becker_evidence_2015,zhu_chasing_2021}, potentially indicate the existence of large neutral IGM islands \citep[e.g.,][]{kulkarni_large_2019,keating_long_2020,nasir_observing_2020,qin_reionization_2021}. This interpretation is further supported by dark gaps in the \lyb\ forest \citep[][]{zhu_long_2022}. Reionization extending to $z<6$ is also consistent with the observed underdensities around long dark gaps traced by \lya\ emitting galaxies \citep[LAEs,][]{becker_evidence_2018,kashino_evidence_2020, christenson_constraints_2021}. In addition, such a late-ending reionization scenario is consistent with the evolution of metal-enriched absorbers at $z\sim6$ \citep[e.g.,][]{becker_evolution_2019,davies_xqr-30_2023,davies_examining_2023-1}, as well as numerical models that reproduce a variety of observations \citep[e.g.,][]{weinberger_modelling_2019,choudhury_studying_2021,qin_reionization_2021,gaikwad_measuring_2023}.

A potentially decisive clue for establishing when reionization ended comes from recent measurements of the mean free path of ionizing photons (\mfp).
\citet[][herein referred to as \citetalias{becker_mean_2021}]{becker_mean_2021} found that the \mfp\ increases by a factor of around ten between $z=6.0$ and 5.1, and the \mfp\ at $z=6.0$ is about eight times shorter than what would be expected based on its evolution at $z \lesssim 5$ (see also constraints from \citealp{bosman_constraints_2021-1}). Such a rapid evolution in the \mfp\ is expected to occur near the end of reionization due to (i) the growth and merger of ionized bubbles, and (ii) the photo-evaporation of dense, optically thick sinks \citep[e.g.,][]{shapiro_photoevaporation_2004, furlanetto_taxing_2005,sobacchi_inhomogeneous_2014,park_hydrodynamic_2016,daloisio_hydrodynamic_2020}. Furthermore, the \mfp\ measurements in \citetalias{becker_mean_2021} are difficult to reconcile with models where reionization completes at $z > 6$, and may instead support models where the IGM is still $\gtrsim 20\%$ neutral at $z=6$ \citep[][]{becker_mean_2021,cain_short_2021,davies_predicament_2021}.

Our understanding of how \mfp\ evolves over $5 < z < 6$ is highly incomplete, however.  The measurements of \citetalias{becker_mean_2021} were restricted to $z \simeq 5.1$ and $ \simeq 6.0$ by a lack of high-quality spectra at intermediate redshifts.  This was due to a historical redshift gap in the discovery of QSOs near $z \sim 5.5$, which have overlap with brown dwarfs in their visible colors.  This gap, however, has been filled by  \citet{yang_discovery_2017,yang_filling_2019} using near- and mid-infrared photometry, making it possible to obtain a significant sample of high-quality $z \sim 5.5$ QSO spectra for the first time.

In this work, we report the first measurements of \mfp\ at multiple redshifts between $z=6$ and 5. In addition to archival QSO spectra used in \citetalias{becker_mean_2021}, our sample includes new QSO spectra from the XQR-30 VLT large program \citep[e.g.,][]{dodorico_xqr-30_2023-1,bischetti_suppression_2022,bosman_hydrogen_2022,chen_measuring_2022,davies_xqr-30_2023,lai_chemical_2022,satyavolu_new_2023-1,zhu_chasing_2021,zhu_long_2022} as well as from new Keck/ESI observations. This paper is organized as follows. In Section~\ref{sec:data}, we describe the data and observations. Section~\ref{sec:methods} briefly introduces the methods we use to measure the \mfp. We present our results and discuss their implications for reionization in Section~\ref{sec:results}. Finally, we summarize our findings in Section~\ref{sec:conslusion}. Throughout this paper we assume a $\Lambda$CDM cosmology with $\Omega_{\rm m}=0.3$, $\Omega_{\Lambda}=0.7$, and $h=0.7$. Distances are quoted in proper units unless otherwise noted.  We also use 912~\AA\ to represent the Lyman limit wavelength of 911.76 \AA.

\section{Data and Observations}\label{sec:data}
To measure \mfp\ over $5<z<6$, we employ a large sample of 97 spectra of QSOs at $4.9<z<6.1$. Our sample includes 23 LRIS spectra and 35 GMOS spectra of QSOs at $z\lesssim 5.3$ used in \citetalias{becker_mean_2021}. For higher redshifts, we use 18 and 6 spectra from the Keck/ESI \citep{sheinis_esi_2002} and VLT/X-Shooter \citep[][]{vernet_x-shooter_2011} archives, respectively. We include 7 high-quality spectra with sufficient wavelength coverage from the XQR-30 VLT large program \citep[][]{dodorico_xqr-30_2023-1}. The rest of the data are new spectra of 8 QSOs near $z\sim 5.5$ from our ESI observations. A summary of our QSO sample is provided in Table \ref{tab:QSOlist}.

\startlongtable
\begin{deluxetable*}{lcclccccr}
    \tablenum{1}
    \tabletypesize{\fontsize{8}{8}\selectfont}
    \tablecaption{QSO Spectra Used for \mfp\ Measurements}
    \tablehead{
        \colhead{\hspace{0.258cm}QSO}\hspace{0.258cm} & 
        \colhead{\hspace{0.258cm} RA (J2000)}\hspace{0.258cm} & 
        \colhead{\hspace{0.258cm} DEC (J2000)}\hspace{0.258cm} & 
        \colhead{\hspace{0.258cm}$z_{\rm qso}$}\hspace{0.258cm} & 
        \colhead{\hspace{0.258cm}Ref}\hspace{0.258cm} & 
        \colhead{\hspace{0.258cm}Instrument} \hspace{0.258cm} & 
        \colhead{\hspace{0.258cm}$M_{1450}$}\hspace{0.258cm} &
        \colhead{\hspace{0.258cm}Ref}\hspace{0.258cm} & 
        \colhead{\hspace{0.258cm} $R_{\rm eq}\,\rm (pMpc)$} \hspace{0.258cm}
    }
    \decimalcolnumbers
    \startdata
        J0015$-$0049      & 00:15:29.86 & $-$00:49:04.3 & $4.931 $ & a & LRIS                 & $-25.2$ & $\alpha$ & $3.8^{+1.1}_{-0.8}$ \\ 
J0256$+$0002      & 02:56:45.75 & $+$00:02:00.2 & $4.960 $ & a & LRIS                 & $-24.6$ & $\alpha$ & $2.8^{+0.9}_{-0.6}$ \\ 
J0236$-$0108      & 02:36:33.83 & $-$01:08:39.2 & $4.974 $ & a & LRIS                 & $-25.0$ & $\alpha$ & $3.4^{+1.1}_{-0.7}$ \\ 
J0338$+$0018      & 03:38:30.02 & $+$00:18:40.0 & $4.988 $ & a & LRIS                 & $-25.1$ & $\alpha$ & $3.5^{+1.2}_{-0.7}$ \\ 
J2226$-$0109      & 22:26:29.28 & $-$01:09:56.6 & $4.994 $ & a & LRIS                 & $-24.6$ & $\alpha$ & $2.8^{+0.9}_{-0.5}$ \\ 
SDSSJ1341$+$4611  & 13:41:41.46 & $+$46:11:10.3 & $5.003 $ & b & GMOS                 & $-25.4$ & $\beta$ & $4.0^{+1.3}_{-0.8}$ \\ 
J0129$-$0028      & 01:29:07.45 & $-$00:28:45.6 & $5.015 $ & a & LRIS                 & $-25.1$ & $\alpha$ & $3.5^{+1.2}_{-0.7}$ \\ 
SDSSJ1337$+$4155  & 13:37:28.82 & $+$41:55:39.9 & $5.018 $ & b & GMOS                 & $-26.6$ & $\beta$ & $7.0^{+2.2}_{-1.4}$ \\ 
J0221$-$0342      & 02:21:12.33 & $-$03:42:31.6 & $5.024 $ & a & LRIS                 & $-24.9$ & $\alpha$ & $3.2^{+1.0}_{-0.7}$ \\ 
SDSSJ0846$+$0800  & 08:46:27.84 & $+$08:00:51.7 & $5.028 $ & b & GMOS                 & $-26.9$ & $\beta$ & $8.1^{+2.6}_{-1.6}$ \\ 
J2111$+$0053      & 21:11:58.02 & $+$00:53:02.6 & $5.034 $ & a & LRIS                 & $-25.3$ & $\alpha$ & $3.9^{+1.2}_{-0.8}$ \\ 
SDSSJ1242$+$5213  & 12:42:47.91 & $+$52:13:06.8 & $5.036 $ & b & GMOS                 & $-25.7$ & $\beta$ & $4.7^{+1.5}_{-1.0}$ \\ 
J0023$-$0018      & 00:23:30.67 & $-$00:18:36.6 & $5.037 $ & a & LRIS                 & $-25.1$ & $\alpha$ & $3.5^{+1.1}_{-0.7}$ \\ 
SDSSJ0338$+$0021  & 03:38:29.31 & $+$00:21:56.2 & $5.040 $ & b & GMOS                 & $-26.7$ & $\beta$ & $7.4^{+2.3}_{-1.5}$ \\ 
J0321$+$0029      & 03:21:55.08 & $+$00:29:41.6 & $5.041 $ & a & LRIS                 & $-24.9$ & $\alpha$ & $3.2^{+1.0}_{-0.7}$ \\ 
SDSSJ0922$+$2653  & 09:22:16.81 & $+$26:53:59.1 & $5.042 $ & b & GMOS                 & $-26.0$ & $\beta$ & $5.4^{+1.6}_{-1.1}$ \\ 
SDSSJ1534$+$1327  & 15:34:59.76 & $+$13:27:01.4 & $5.043 $ & b & GMOS                 & $-25.0$ & $\beta$ & $3.4^{+1.0}_{-0.7}$ \\ 
SDSSJ1101$+$0531  & 11:01:34.36 & $+$05:31:33.9 & $5.045 $ & b & GMOS                 & $-27.7$ & $\beta$ & $11.8^{+3.6}_{-2.5}$ \\ 
SDSSJ1340$+$3926  & 13:40:15.04 & $+$39:26:30.8 & $5.048 $ & b & GMOS                 & $-26.8$ & $\beta$ & $7.8^{+2.4}_{-1.6}$ \\ 
SDSSJ1423$+$1303  & 14:23:25.92 & $+$13:03:00.7 & $5.048 $ & b & GMOS                 & $-27.1$ & $\beta$ & $8.9^{+2.8}_{-1.8}$ \\ 
SDSSJ1154$+$1341  & 11:54:24.73 & $+$13:41:45.8 & $5.060 $ & b & GMOS                 & $-25.6$ & $\beta$ & $4.5^{+1.3}_{-1.0}$ \\ 
J1408$+$5300      & 14:08:22.92 & $+$53:00:20.9 & $5.072 $ & a & LRIS                 & $-25.5$ & $\alpha$ & $4.3^{+1.3}_{-0.9}$ \\ 
SDSSJ1614$+$2059  & 16:14:47.04 & $+$20:59:02.8 & $5.081 $ & b & GMOS                 & $-26.6$ & $\beta$ & $7.2^{+2.0}_{-1.6}$ \\ 
J2312$+$0100      & 23:12:16.44 & $+$01:00:51.6 & $5.082 $ & a & LRIS                 & $-25.6$ & $\alpha$ & $4.5^{+1.3}_{-1.0}$ \\ 
J2239$+$0030      & 22:39:07.56 & $+$00:30:22.5 & $5.092 $ & a & LRIS                 & $-25.2$ & $\alpha$ & $3.8^{+1.0}_{-0.9}$ \\ 
SDSSJ1204$-$0021  & 12:04:41.73 & $-$00:21:49.5 & $5.094 $ & b & GMOS                 & $-27.4$ & $\beta$ & $10.5^{+2.9}_{-2.4}$ \\ 
J2233$-$0107      & 22:33:27.65 & $-$01:07:04.5 & $5.104 $ & a & LRIS                 & $-25.0$ & $\alpha$ & $3.5^{+1.0}_{-0.8}$ \\ 
J0108$-$0100      & 01:08:29.97 & $-$01:00:15.7 & $5.118 $ & a & LRIS                 & $-24.6$ & $\alpha$ & $2.9^{+0.8}_{-0.7}$ \\ 
SDSSJ1222$+$1958  & 12:22:37.96 & $+$19:58:42.9 & $5.120 $ & b & GMOS                 & $-25.5$ & $\beta$ & $4.4^{+1.2}_{-1.0}$ \\ 
SDSSJ0913$+$5919  & 09:13:16.55 & $+$59:19:21.7 & $5.122 $ & b & GMOS                 & $-25.3$ & $\beta$ & $4.0^{+1.1}_{-0.9}$ \\ 
SDSSJ1209$+$1831  & 12:09:52.71 & $+$18:31:47.0 & $5.127 $ & b & GMOS                 & $-26.8$ & $\beta$ & $8.0^{+2.2}_{-1.8}$ \\ 
SDSSJ1148$+$3020  & 11:48:26.17 & $+$30:20:19.3 & $5.128 $ & b & GMOS                 & $-26.3$ & $\beta$ & $6.3^{+1.7}_{-1.5}$ \\ 
SDSSJ1334$+$1220  & 13:34:12.56 & $+$12:20:20.7 & $5.130 $ & b & GMOS                 & $-26.8$ & $\beta$ & $8.0^{+2.2}_{-1.9}$ \\ 
J2334$-$0010      & 23:34:55.07 & $-$00:10:22.2 & $5.137 $ & a & LRIS                 & $-24.6$ & $\alpha$ & $2.9^{+0.8}_{-0.7}$ \\ 
J0115$+$0015      & 01:15:44.78 & $+$00:15:15.0 & $5.144 $ & a & LRIS                 & $-25.1$ & $\alpha$ & $3.7^{+1.0}_{-0.9}$ \\ 
SDSSJ2228$-$0757  & 22:28:45.14 & $-$07:57:55.3 & $5.150 $ & b & GMOS                 & $-26.1$ & $\beta$ & $5.8^{+1.6}_{-1.3}$ \\ 
SDSSJ1050$+$5804  & 10:50:36.47 & $+$58:04:24.6 & $5.151 $ & b & GMOS                 & $-26.5$ & $\beta$ & $7.0^{+1.9}_{-1.6}$ \\ 
SDSSJ1054$+$1633  & 10:54:45.43 & $+$16:33:37.4 & $5.154 $ & b & GMOS                 & $-26.4$ & $\beta$ & $6.6^{+1.8}_{-1.5}$ \\ 
SDSSJ0957$+$0610  & 09:57:07.67 & $+$06:10:59.6 & $5.167 $ & b & GMOS                 & $-27.6$ & $\beta$ & $11.6^{+3.1}_{-2.7}$ \\ 
J2238$-$0027      & 22:38:50.19 & $-$00:27:01.8 & $5.172 $ & a & LRIS                 & $-25.1$ & $\alpha$ & $3.7^{+1.0}_{-0.8}$ \\ 
SDSSJ0854$+$2056  & 08:54:30.37 & $+$20:56:50.9 & $5.179 $ & b & GMOS                 & $-27.0$ & $\beta$ & $8.8^{+2.4}_{-2.1}$ \\ 
SDSSJ1132$+$1209  & 11:32:46.50 & $+$12:09:01.7 & $5.180 $ & b & GMOS                 & $-27.2$ & $\beta$ & $9.6^{+2.5}_{-2.2}$ \\ 
J1414$+$5732      & 14:14:31.57 & $+$57:32:34.1 & $5.188 $ & a & LRIS                 & $-24.8$ & $\alpha$ & $3.2^{+0.9}_{-0.7}$ \\ 
SDSSJ0915$+$4924  & 09:15:43.64 & $+$49:24:16.6 & $5.199 $ & b & GMOS                 & $-26.9$ & $\beta$ & $8.4^{+2.2}_{-1.9}$ \\ 
SDSSJ1221$+$4445  & 12:21:46.42 & $+$44:45:28.0 & $5.203 $ & b & GMOS                 & $-25.8$ & $\beta$ & $5.1^{+2.3}_{-0.7}$ \\ 
SDSSJ0824$+$1302  & 08:24:54.01 & $+$13:02:17.0 & $5.207 $ & b & GMOS                 & $-26.2$ & $\beta$ & $6.1^{+2.7}_{-0.8}$ \\ 
J0349$+$0034      & 03:49:59.42 & $+$00:34:03.5 & $5.209 $ & a & LRIS                 & $-25.3$ & $\alpha$ & $4.1^{+1.8}_{-0.5}$ \\ 
SDSSJ0902$+$0851  & 09:02:45.76 & $+$08:51:15.9 & $5.226 $ & b & GMOS                 & $-25.9$ & $\beta$ & $5.4^{+2.3}_{-0.8}$ \\ 
SDSSJ1436$+$2132  & 14:36:05.00 & $+$21:32:39.2 & $5.227 $ & b & GMOS                 & $-26.8$ & $\beta$ & $8.2^{+3.3}_{-1.2}$ \\ 
J2202$+$0131      & 22:02:33.20 & $+$01:31:20.3 & $5.229 $ & a & LRIS                 & $-24.6$ & $\alpha$ & $3.0^{+1.2}_{-0.5}$ \\ 
J0208$-$0112      & 02:08:04.31 & $-$01:12:34.4 & $5.231 $ & a & LRIS                 & $-25.3$ & $\alpha$ & $4.1^{+1.7}_{-0.6}$ \\ 
J2211$+$0011      & 22:11:41.02 & $+$00:11:19.0 & $5.237 $ & a & LRIS                 & $-24.8$ & $\alpha$ & $3.3^{+1.3}_{-0.6}$ \\ 
SDSSJ1053$+$5804  & 10:53:22.98 & $+$58:04:12.1 & $5.250 $ & b & GMOS                 & $-27.0$ & $\beta$ & $9.3^{+3.5}_{-1.6}$ \\ 
SDSSJ1341$+$3510  & 13:41:54.02 & $+$35:10:05.8 & $5.252 $ & b & GMOS                 & $-26.6$ & $\beta$ & $7.7^{+2.9}_{-1.4}$ \\ 
SDSSJ1026$+$2542  & 10:26:23.62 & $+$25:42:59.4 & $5.254 $ & b & GMOS                 & $-26.5$ & $\beta$ & $7.4^{+2.7}_{-1.3}$ \\ 
SDSSJ1626$+$2751  & 16:26:26.50 & $+$27:51:32.5 & $5.265 $ & b & GMOS                 & $-27.8$ & $\beta$ & $13.6^{+4.7}_{-2.5}$ \\ 
SDSSJ1202$+$3235  & 12:02:07.78 & $+$32:35:38.8 & $5.298 $ & a & ESI                  & $-28.0$ & $\beta$ & $15.8^{+4.7}_{-3.5}$ \\ 
SDSSJ1233$+$0622  & 12:33:33.47 & $+$06:22:34.2 & $5.300 $ & b & GMOS                 & $-26.2$ & $\beta$ & $6.8^{+2.0}_{-1.5}$ \\ 
SDSSJ1614$+$4640  & 16:14:25.13 & $+$46:40:28.9 & $5.313 $ & b & GMOS                 & $-25.8$ & $\beta$ & $5.7^{+1.6}_{-1.2}$ \\ 
SDSSJ1659$+$2709  & 16:59:02.12 & $+$27:09:35.1 & $5.316 $ & a & ESI                  & $-27.9$ & $\beta$ & $14.6^{+4.2}_{-3.2}$ \\ 
SDSSJ1437$+$2323  & 14:37:51.82 & $+$23:23:13.4 & $5.320 $ & a & ESI                  & $-26.8$ & $\beta$ & $9.0^{+2.6}_{-2.1}$ \\ 
J1656$+$4541      & 16:56:35.46 & $+$45:41:13.5 & $5.336 $ & a & ESI                  & $-27.6$ & $\gamma$ & $12.9^{+3.6}_{-2.9}$ \\ 
SDSSJ1340$+$2813  & 13:40:40.24 & $+$28:13:28.1 & $5.349 $ & a & ESI                  & $-26.6$ & $\beta$ & $8.3^{+2.3}_{-2.0}$ \\ 
J0306$+$1853      & 03:06:42.51 & $+$18:53:15.8 & $5.3808$ & c & ESI$^\ddagger$       & $-28.9$ & $\gamma$ & $24.5^{+6.6}_{-5.9}$ \\ 
J0155$+$0415      & 01:55:33.28 & $+$04:15:06.7 & $5.382 $ & a & ESI                  & $-27.0$ & $\delta$ & $10.3^{+2.7}_{-2.5}$ \\ 
SDSSJ0231$-$0728  & 02:31:37.65 & $-$07:28:54.5 & $5.420 $ & a & ESI                  & $-26.6$ & $\beta$ & $8.6^{+2.0}_{-2.2}$ \\ 
J1054$+$4637      & 10:54:05.32 & $+$46:37:30.2 & $5.469 $ & a & ESI$^\ddagger$       & $-27.0$ & $\delta$ & $10.3^{+2.3}_{-2.8}$ \\ 
SDSSJ1022$+$2252  & 10:22:10.04 & $+$22:52:25.4 & $5.4787$ & c & ESI                  & $-27.3$ & $\epsilon$ & $11.9^{+2.4}_{-3.2}$ \\ 
J1513$+$0854      & 15:13:39.64 & $+$08:54:06.5 & $5.4805$ & c & ESI$^\ddagger$       & $-26.8$ & $\delta$ & $9.6^{+2.0}_{-2.7}$ \\ 
J0012$+$3632      & 00:12:32.88 & $+$36:32:16.1 & $5.485 $ & c & ESI$^\ddagger$       & $-27.2$ & $\delta$ & $11.8^{+2.4}_{-3.3}$ \\ 
J2207$-$0416      & 22:07:10.12 & $-$04:16:56.3 & $5.5297$ & c & ESI$^\ddagger$       & $-27.8$ & $\delta$ & $15.3^{+7.9}_{-2.6}$ \\ 
J2317$+$2244      & 23:17:38.25 & $+$22:44:09.6 & $5.5580$ & c & ESI$^\ddagger$       & $-27.4$ & $\delta$ & $12.9^{+6.5}_{-2.2}$ \\ 
J1500$+$2816      & 15:00:36.84 & $+$28:16:03.0 & $5.5727$ & c & ESI$^\ddagger$       & $-27.6$ & $\delta$ & $14.4^{+6.9}_{-2.6}$ \\ 
J1650$+$1617      & 16:50:42.26 & $+$16:17:21.5 & $5.5769$ & c & ESI$^\ddagger$       & $-27.2$ & $\delta$ & $12.3^{+5.8}_{-2.3}$ \\ 
J0108$+$0711      & 01:08:06.59 & $+$07:11:20.7 & $5.580 $ & a & ESI$^\ddagger$       & $-27.2$ & $\delta$ & $11.9^{+5.7}_{-2.2}$ \\ 
J1335$-$0328      & 13:35:56.24 & $-$03:28:38.3 & $5.699 $ & c & X-Shooter             & $-27.7$ & $\delta$ & $18.0^{+4.1}_{-5.8}$ \\ 
SDSSJ0927$+$2001  & 09:27:21.82 & $+$20:01:23.6 & $5.7722$ & d & X-Shooter             & $-26.8$ & $\zeta$ & $12.8^{+1.6}_{-4.8}$ \\ 
SDSSJ1044$-$0125  & 10:44:33.04 & $-$01:25:02.2 & $5.7847$ & e & ESI                  & $-27.2$ & $\eta$ & $15.6^{+1.8}_{-5.9}$ \\ 
PSOJ308$-$27      & 20:33:55.91 & $-$27:38:54.6 & $5.798 $ & a & X-Shooter$^\dagger$  & $-26.8$ & $\zeta$ & $13.3^{+1.3}_{-5.3}$ \\ 
SDSSJ0836$+$0054  & 08:36:43.85 & $+$00:54:53.3 & $5.805 $ & a & ESI                  & $-27.8$ & $\zeta$ & $21.4^{+12.1}_{-5.0}$ \\ 
SDSSJ0002$+$2550  & 00:02:39.40 & $+$25:50:34.8 & $5.824 $ & a & ESI                  & $-27.3$ & $\zeta$ & $17.3^{+9.2}_{-4.2}$ \\ 
PSOJ065$+$01      & 04:23:50.15 & $+$01:43:24.8 & $5.8348$ & f & X-Shooter$^\dagger$  & $-26.6$ & $\zeta$ & $12.4^{+6.3}_{-3.1}$ \\ 
PSOJ025$-$11      & 01:40:57.03 & $-$11:40:59.5 & $5.8414$ & f & X-Shooter$^\dagger$  & $-26.9$ & $\zeta$ & $14.4^{+7.4}_{-3.7}$ \\ 
SDSSJ0840$+$5624  & 08:40:35.09 & $+$56:24:19.8 & $5.8441$ & g & ESI                  & $-27.2$ & $\zeta$ & $16.8^{+8.4}_{-4.4}$ \\ 
PSOJ242$-$12      & 16:09:45.53 & $-$12:58:54.1 & $5.8468$ & f & X-Shooter$^\dagger$  & $-26.9$ & $\zeta$ & $14.8^{+7.4}_{-3.9}$ \\ 
PSOJ023$-$02      & 01:32:01.70 & $-$02:16:03.1 & $5.848 $ & a & X-Shooter$^\dagger$  & $-26.5$ & $\zeta$ & $12.0^{+6.1}_{-3.0}$ \\ 
SDSSJ0005$-$0006  & 00:05:52.33 & $-$00:06:55.6 & $5.851 $ & a & ESI                  & $-25.7$ & $\zeta$ & $8.4^{+4.1}_{-2.2}$ \\ 
PSOJ183$-$12      & 12:13:11.81 & $-$12:46:03.5 & $5.899 $ & a & X-Shooter$^\dagger$  & $-27.5$ & $\zeta$ & $20.1^{+8.8}_{-5.8}$ \\ 
SDSSJ1411$+$1217  & 14:11:11.28 & $+$12:17:37.3 & $5.920 $ & a & ESI                  & $-26.7$ & $\zeta$ & $14.0^{+5.8}_{-4.1}$ \\ 
PSOJ340$-$18      & 22:40:48.98 & $-$18:39:43.8 & $6.0007$ & h & X-Shooter             & $-26.4$ & $\zeta$ & $12.4^{+5.1}_{-3.8}$ \\ 
SDSSJ0818$+$1722  & 08:18:27.40 & $+$17:22:52.0 & $6.001 $ & a & X-Shooter             & $-27.5$ & $\zeta$ & $20.6^{+8.2}_{-6.4}$ \\ 
SDSSJ1137$+$3549  & 11:37:17.72 & $+$35:49:56.9 & $6.030 $ & a & ESI                  & $-27.4$ & $\zeta$ & $19.7^{+8.0}_{-6.0}$ \\ 
SDSSJ1306$+$0356  & 13:06:08.26 & $+$03:56:26.2 & $6.0330$ & i & X-Shooter             & $-26.8$ & $\zeta$ & $14.9^{+6.1}_{-4.6}$ \\ 
ULASJ1207$+$0630  & 12:07:37.43 & $+$06:30:10.1 & $6.0366$ & j & X-Shooter             & $-26.6$ & $\zeta$ & $13.6^{+5.4}_{-4.3}$ \\ 
SDSSJ2054$-$0005  & 20:54:06.49 & $-$00:05:14.6 & $6.0391$ & e & ESI                  & $-26.2$ & $\zeta$ & $11.3^{+4.6}_{-3.5}$ \\ 
SDSSJ0842$+$1218  & 08:42:29.43 & $+$12:18:50.6 & $6.0763$ & j & X-Shooter$^\dagger$  & $-26.9$ & $\zeta$ & $15.6^{+6.2}_{-4.9}$ \\ 
SDSSJ1602$+$4228  & 16:02:53.98 & $+$42:28:24.9 & $6.084 $ & a & ESI                  & $-26.9$ & $\zeta$ & $15.6^{+6.2}_{-4.7}$ \\ 

    \enddata
    \tablecomments{Columns: (1) QSO name; 
    (2 \& 3) QSO coordinates; 
    (4) QSO redshift; 
    (5) reference for QSO redshift; 
    (6) instrument used for \mfp\ measurements: $\dagger$ and $\ddagger$ denote XQR-30 spectra and spectra from our new ESI observations, respectively;
    (7) absolute magnitude corresponding to the mean luminosity at rest-frame 1450 \AA;
    (8) reference for $M_{1450}$;
    (9) QSO proximity zone size defined by Equation \ref{eq:req}.
    }

    \tablerefs{Redshift lines and references: 
    a. apparent start of the \lya\ forest: \citet{becker_evolution_2019,becker_mean_2021} and updated measurements in this work (after correction, see text);
    b. adopted from \citet[][]{worseck_giant_2014};
    c. [\ion{C}{2}] 158$\mu$m: this work;
    d. CO: \citet{carilli_detection_2007};
    e. [\ion{C}{2}] 158$\mu$m: \citet{wang_star_2013};
    f. [\ion{C}{2}] 158$\mu$m: Bosman et al.~(in prep.);
    g. CO: \citet{wang_molecular_2010};
    h. \lya\ halo: \citet{farina_requiem_2019};
    i. [\ion{C}{2}] 158$\mu$m: \citet{venemans_kiloparsec-scale_2020};
    j. [\ion{C}{2}] 158$\mu$m: \citet{decarli_alma_2018};.
    \\
    $M_{1450}$ references: 
    $\alpha$. \citet{mcgreer_z_2013,mcgreer_faint_2018};
    $\beta$. \citet[][]{becker_mean_2021}, in which $M_{1450}$ values are calculated from the flux-calibrated spectra published by \citet{worseck_giant_2014};
    $\gamma$. \citet{wang_survey_2016};
    $\delta$. \citet{yang_discovery_2017,yang_filling_2019};
    $\epsilon$. measured in this work;    
    $\zeta$. \citet[][]{banados_pan-starrs1_2016,banados_pan-starrs1_2023};
    $\eta$. \citet{schindler_x-shooteralma_2020-1}.
    }
    \label{tab:QSOlist}
\end{deluxetable*}

\begin{figure*}
    \centering \includegraphics[width=6in]{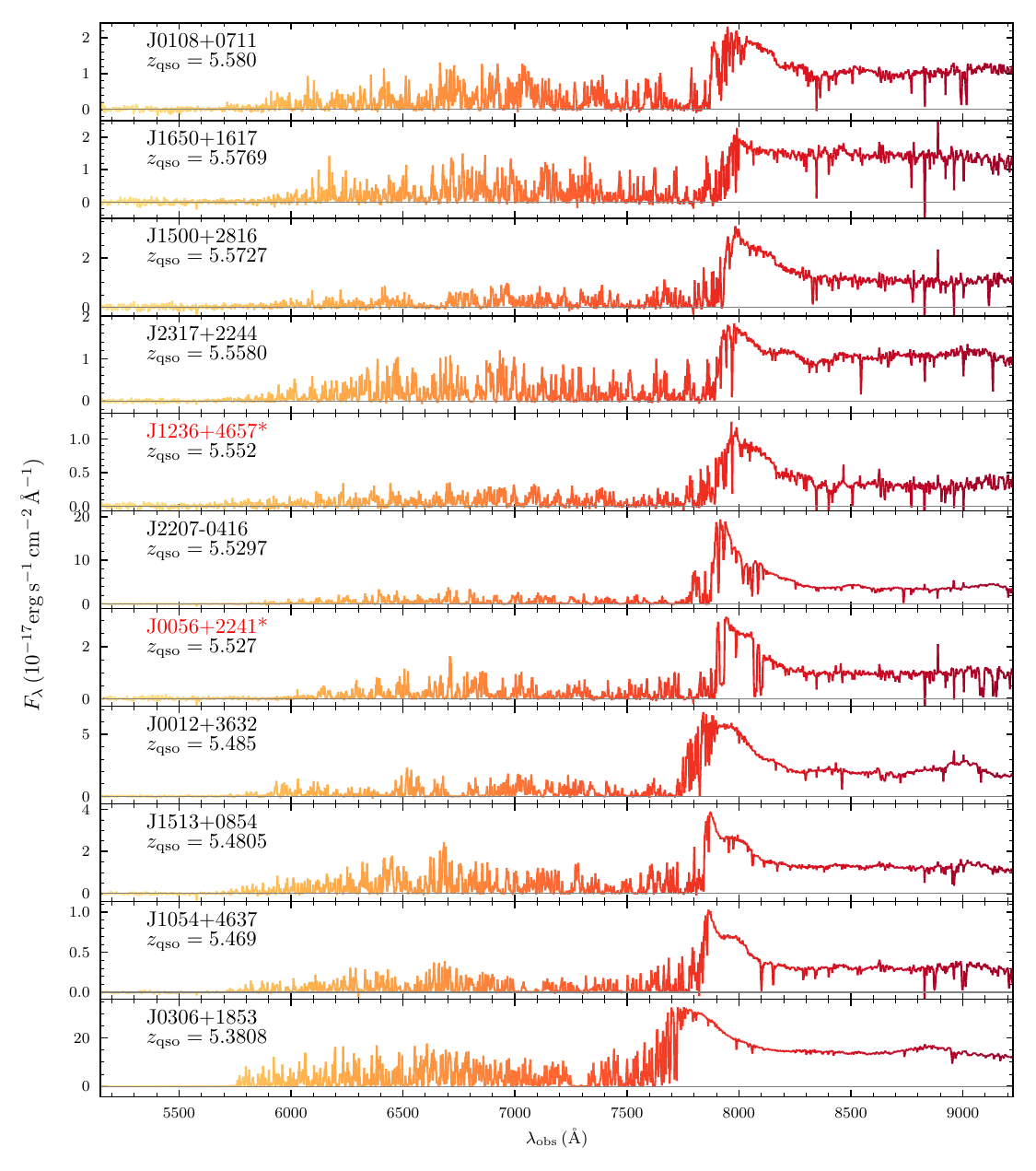}
    \caption{New Keck/ESI QSO spectra obtained for this work.  Orange-red and gray curves plot the flux and zero flux, respectively. J0056$+$2241 and J1236$+$4657 (labelled with *) are not included in our sample of \mfp\ measurements (see text for details). The spectra are re-binned to 2~\AA\ for display.  
    }
    \label{fig:spectra}
\end{figure*}

\begin{figure*}
    \centering
    \includegraphics[width=5.9in]{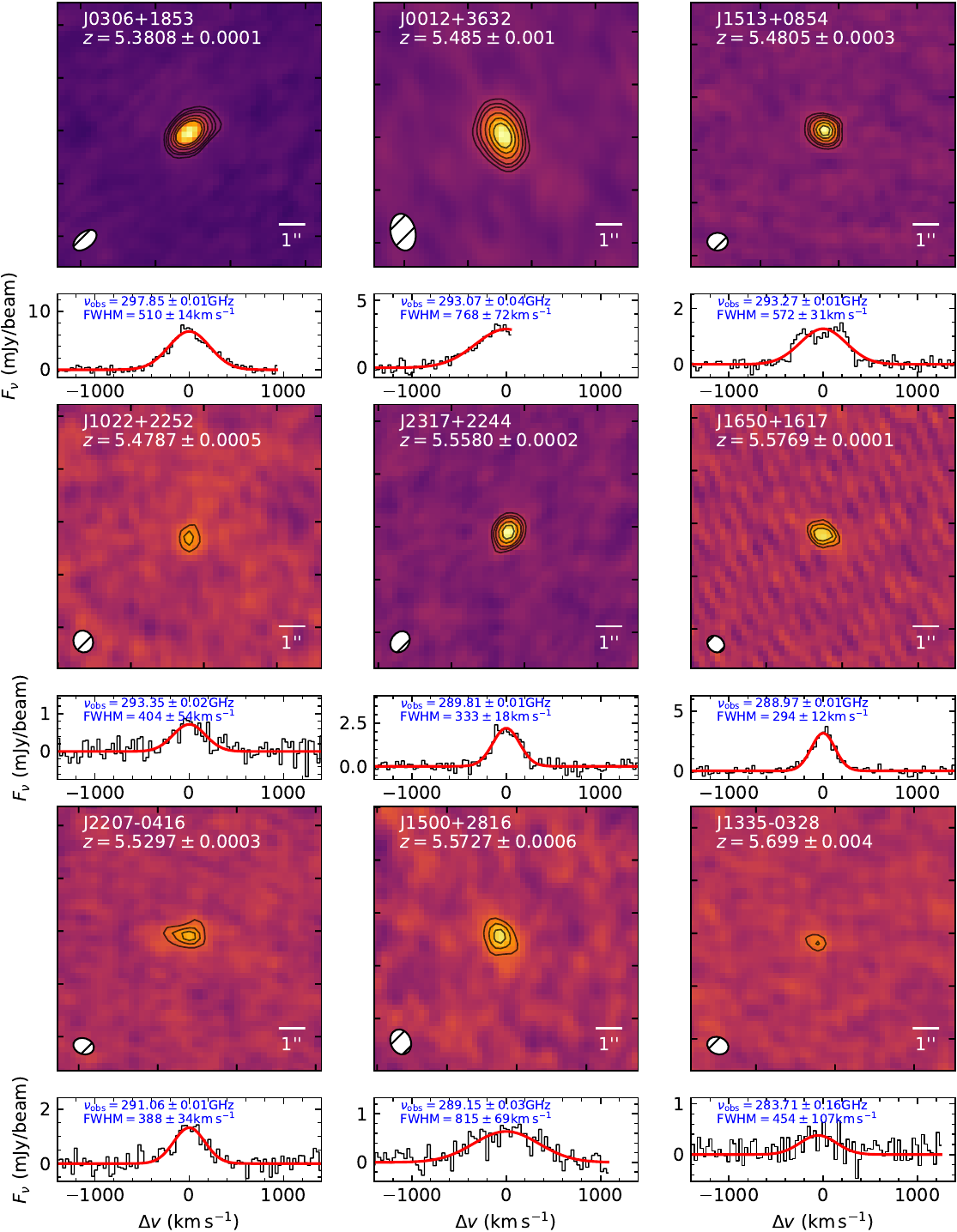}
    \caption{[\ion{C}{2}] 158$\mu$m emission maps and spectra of QSOs with our new Keck/ESI observations. Contours show (2$\sigma$, 3$\sigma$, 4$\sigma$, 6$\sigma$, 8$\sigma$, 10$\sigma$) levels. Measured redshifts are labelled for each QSO and red curves show the best Gaussian fits. Observed frequency and FWHM of the [\ion{C}{2}] emission are also provided for reference.}
    \label{fig:cii}
\end{figure*}

In 2021 and 2022, we targeted bright QSOs with $z$-band magnitude $m_z\le 20$ presented in \citet{yang_discovery_2017,yang_filling_2019} using ESI. With a typical exposure time of $\sim 1-3$ hours and using the $1.00''$ and $0.75''$ slits, we acquired spectra for 11 objects (Figure \ref{fig:spectra}), including 10 QSOs at $z_{\rm qso}\sim 5.5$ (8 of them are included in our sample; see below) and 1 QSO at a lower redshift for replacing its archival spectrum. We followed \citet{becker_evolution_2019} to reduce the data, using a custom pipeline that includes optimal techniques for sky subtraction \citep{kelson_optimal_2003}, one-dimensional spectral extraction \citep{horne_optimal_1986}, and telluric absorption corrections for individual exposures using models based on the atmospheric conditions measured by the Cerro Paranal Advanced Sky Model \citep{noll_atmospheric_2012,jones_advanced_2013}. We extracted the spectra with a pixel size of 15 \kms.  The typical resolution full width at half maximum (FWHM) is approximately 45 \kms.

All targets in our sample were selected without any foreknowledge of the Lyman continuum (LyC) transmission. We include all usable spectra as long as the QSO is free from very strong associated metal absorption and/or associated \lya\ damping wing absorption, which may bias the \mfp\ measurements. We also reject objects with strong broad absorption lines (BALs) near the systemic redshift (see \citealp{bischetti_suppression_2022} for the XQR-30 spectra). Because the LyC transmission is very weak at $z>5.3$, we only use spectra with a signal-to-noise ratio (SNR) of $\gtrsim 20$ per 30 \kms\ near 1285~\AA\ in the rest frame. Among the objects with new ESI spectra, we exclude J0056 due to its strong associated absorber, and J1236 for its low SNR.

For QSO redshifts, we employ measurements based on sub-millimeter observations, whenever available. Additionally, we carried out ALMA Band 7 observations of our new ESI targets in Cycle 9 and determined the systemic redshifts by fitting the [\ion{C}{2}] 158$\mu$m line. For each object, we used two overlapping spectral windows to cover the [\ion{C}{2}] line based on the estimated redshift and another two spectral windows to cover the dust continuum. With C43-(1, 2, 3) configurations, the typical angular resolution is $\sim 1''$. The data are calibrated and reduced using the default procedures of the CASA pipeline (version 6.4.1.12; \citealp{mcmullin_casa_2007,casa_team_casa_2022}). We follow the procedures described in \citet{eilers_detecting_2020} to generate the data cube and image the [\ion{C}{2}] line: the [\ion{C}{2}] emission is continuum subtracted with {\tt uvconstsub}, and imaged with the {\tt tclean} procedure using Briggs cleaning and a robust parameter of 2 (natural weighting) to maximize the sensitivity. We use a robust parameter of 0.5 for J1650+1617 to achieve a best data product. The mean RMS noise of our data set is 0.25 ${\rm mJy\,beam^{-1}}$ per 30 MHz bin. Figure \ref{fig:cii} displays [\ion{C}{2}] maps along with Gaussian fits to the emission. For each QSO, we extract the spectrum within one beam size centered at the target. We create the [\ion{C}{2}] map by stacking the data cube within 1 standard deviation of the Gaussian fit from the line center. We note that the emission line of J0012$+$3632 is incomplete because the [\ion{C}{2}] line is at the edge of our spectral window, which was chosen based on a preliminary redshift estimate \citep[][]{yang_filling_2019}. J1513$+$0854 shows a double-peak emission line, which may be due to the rotating disk of the QSO host galaxy. Future observations with higher spatial resolution may help resolve the disk.

\begin{figure*}
    \centering \includegraphics[width=6in]{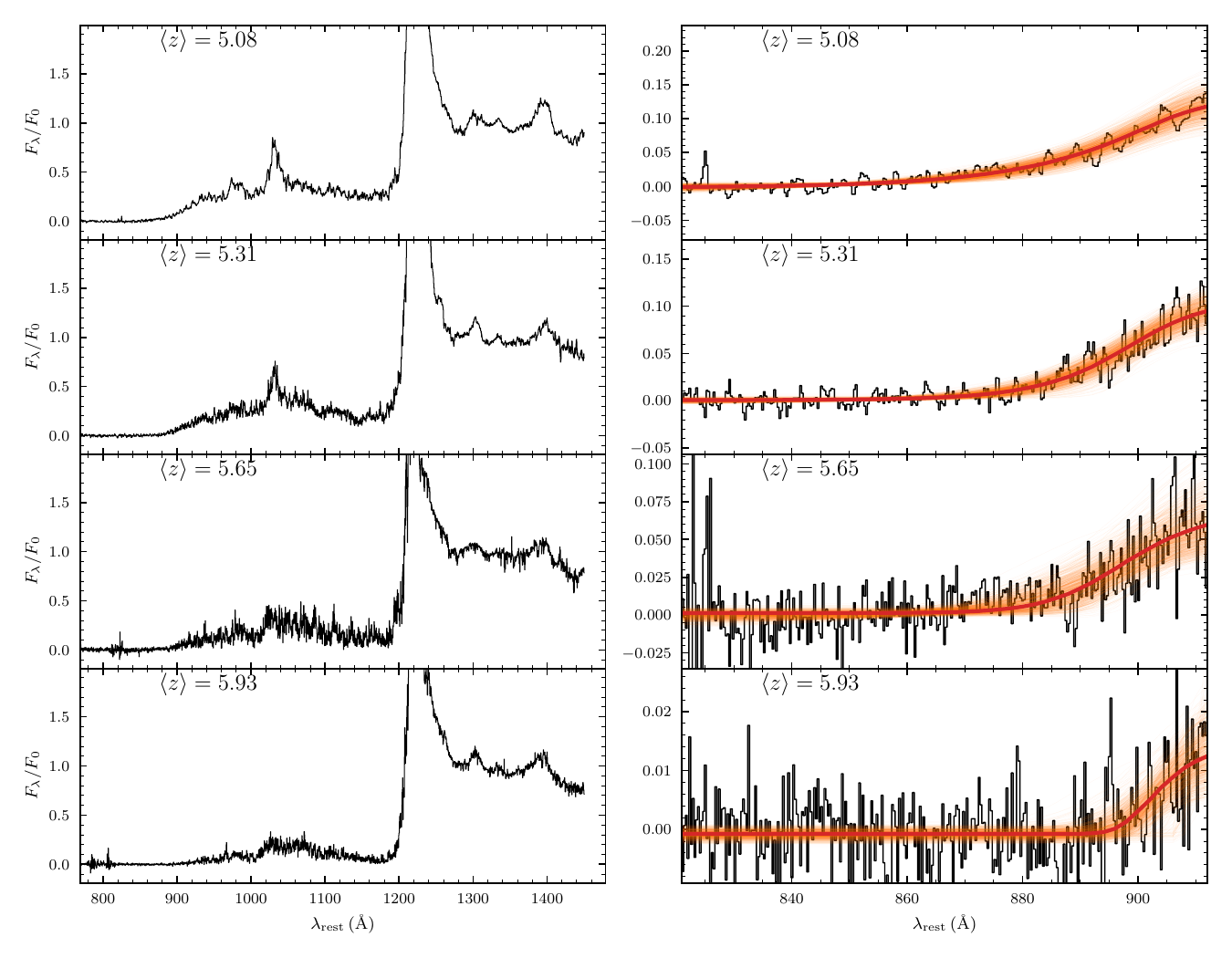}
    \caption{Composite QSO spectra for each redshift bin (left-hand panel) and fits to the LyC transmission (right-hand panel). Black curves show observed flux normalized by the median flux over 1270-1380~\AA\ in the rest frame. The red curve shows our best-fit model.  Thin orange curves show the fits from bootstrap realization (only 1000 curves are plotted here for display).}
    \label{fig:composites}
\end{figure*}

For the rest of our sample, we employ redshifts measured from the apparent start of the \lya\ forest, which are determined for each line of sight by visually searching for the first \lya\ absorption line blueward the \lya\ peak \citep[e.g.,][]{worseck_giant_2014,becker_evolution_2019}. We do not use redshifts measured from \ion{Mg}{2} emission because of their large offsets ($\sim$ 500 \kms) from the systemic redshifts \citep[e.g.,][]{venemans_bright_2016,mazzucchelli_physical_2017,schindler_x-shooteralma_2020-1}. Based on 42 QSOs at $5.3<z_{\rm qso}<6.6$ with [\ion{C}{2}] or CO redshifts, we find that the redshifts we measure from the apparent start of the \lya\ forest are blueshifted from the [\ion{C}{2}] or CO redshifts by $\sim185$ \kms\ on average, with a standard deviation of $\sim370$ \kms. Such a redshift offset can be explained by the strong proximity zone effect close to the QSO: the first significant absorber may typically occur slightly blueward of the QSO redshift due to ionization effects. This offset is also consistent with that measured in \citetalias{becker_mean_2021}. Thus, we shift redshifts measured from the apparent start of the \lya\ forest by $+185$ \kms\ when measuring \mfp, and the corrected values are listed in Table \ref{tab:QSOlist}.
We generate rest-frame composite spectra for QSOs in each $\Delta z=0.3$ bin, starting from $z=4.9$. The redshift bins with a mean redshift of $\langle z \rangle = 5.08$, 5.31, 5.65, and 5.93 consist of 44, 26, 9, and 18 spectra, respectively. 
Following \citetalias{becker_mean_2021}, each spectrum is shifted to rest-frame wavelength before being normalized. The normalization is done by dividing each spectrum by its \textit{continuum} flux measured over wavelengths in the rest frame  where the flux from broad emission lines is minimal. Here, we use the \textit{continuum} flux over $1270 {\rm \AA}< \lambda < 1380 {\rm \AA}$ in the rest frame. We have tested that using a different wavelength window does not significantly change our results.
Additionally, we identify and mask wavelength regions affected by skyline subtraction residuals. To reject spurious bad pixels, we apply a light median filter using a 3-pixel sliding window. Mean composite spectra are then computed in 120 \kms\ bins (as shown in the left-hand panel of Figure \ref{fig:composites}).

\section{Methods}\label{sec:methods}

We measure the \mfp, which is defined as the distance travelled by ionizing photons that would be attenuated by a factor of $1/e$ by LyC absorption, by fitting the transmitted flux profile blueward of the Lyman limit in composite QSO spectra \citepalias[\citealp{prochaska_direct_2009};][]{becker_mean_2021}.  One challenge with this approach is that the LyC transmission at $z>5$ can be significantly affected by the QSO proximity effect.  The ionizing flux from the QSO decreases the ionizing opacity in the proximity zone, which can bias the inferred \mfp\ high by a factor of two or more \citepalias[\citealp{daloisio_large_2018};][]{becker_mean_2021}.  This is especially important when \mfp\ is smaller than the proximity zone size, which is true for bright QSOs at $z \gtrsim 5$. 

To address this bias, we follow the methods and modeling presented in \citetalias[][]{becker_mean_2021}, which modified the \citet[][]{prochaska_direct_2009} method of measuring \mfp\ to explicitly include the proximity effect. Motivated by simulations, we account for the decrease in ionizing opacity near the QSO by scaling the opacity, $\kappa_{\rm LL}$, according to the local photoionization rate, $\Gamma$.  This dependence is modeled as a power law such that $\kappa_{\rm LL} \propto \Gamma^{-\xi}$,
\begin{equation}\label{eq:kappa}
    \kappa_{\rm LL} (r) = \kappa_{\rm LL}^{\rm bg} \left[  1+\frac{\Gamma_{\rm qso}(r)}{\Gamma_{\rm bg}} \right]^{-\xi} ~,
\end{equation}
where $\Gamma_{\rm qso}(r)$ is the photoionization rate due to the QSO at a distance $r$ from the QSO, and $\Gamma_{\rm bg}$ is the background photoionization rate.
In order to calculate $\Gamma_{\rm qso}(r)$, a key parameter used to describe the proximity zone effect in \citetalias[][]{becker_mean_2021} is $R_{\rm eq}$. It denotes the distance from the QSO where $\Gamma_{\rm qso}(r)$ and $\Gamma_{\rm bg}$ would be equal for purely geometric dilution in the absence of attenuation.
Following \citet{calverley_measurements_2011}, $R_{\rm eq}$ is related to $\Gamma_{\rm bg}$ and the ionizing luminosity of the QSO, $L_{912}$, by
\begin{equation}\label{eq:req}
    R_{\rm eq} = \left[ \frac{L_{912} \sigma_0}{8\pi^2 \hbar \Gamma_{\rm bg}(\alpha_{\nu}^{\rm ion} + 2.75)}  \right]^{1/2} ~,
\end{equation}
where $\sigma_0$ and $\alpha_{\nu}^{\rm ion}$ are the \HI\ ionization cross section at 912~\AA\ and the power-law index of the QSO continuum at $\lambda<912$~\AA\ in the frequency domain, respectively, and $\hbar$ is the reduced Planck constant. $L_{912}$ can be further related to the absolute magnitude corresponding the the mean luminosity at rest-frame 1450 \AA, $L_{1450}$, by $L_{912}=L_{1450}(\nu_{912}/\nu_{1450})^{-\alpha_{\nu}^{\rm UV}}$. Here, $\nu_{912}$ and $\nu_{1450}$ are the photon frequencies at 912~\AA\ and 1450 \AA, respectively, and $\alpha_{\nu}^{\rm UV}$ is the power-law slope for the non-ionizing ($912\rm \AA < \lambda < 1450 \AA$) continuum of the QSO continuum. Following \citetalias{becker_mean_2021}, we adopt $\alpha_{\nu}^{\rm ion} = 1.5\pm0.3$ \footnote{$\alpha_{\lambda}^{\rm ion} = -0.5\pm0.3$ in the wavelength domain.} \citep{telfer_rest-frame_2002,stevans_hst-cos_2014,lusso_first_2015} and $\alpha_{\nu}^{\rm UV} = 0.6 \pm 0.1$ \citep[][see also \citealp{vanden_berk_composite_2001,shull_hst-cos_2012,stevans_hst-cos_2014}]{lusso_first_2015}. 
Following \citetalias[][]{becker_mean_2021}, the observed flux, $f_{\lambda}^{\rm obs}$, is given by the mean intrinsic QSO continuum, $f_{\lambda}^{\rm cont} \propto (\lambda/912\rm \AA)^{\alpha_{\lambda}^{\rm ion}}$,  attenuated by the effective opacity of the Lyman series in the foreground IGM, and the LyC optical depth. The foreground Lyman series opacity is calculated by
\begin{equation}
    \tau_{\rm eff}^{\rm Lyman} (\lambda_{\rm obs}) = \sum_{j=\rm Ly\alpha,\, Ly\beta,\, ...,\, Ly40}\teff^j(z_j)\, ,    
\end{equation}
where $\teff^j(z_j)$ is the effective opacity of transition $j$ at redshift $z_j$ such that $(1+z_j)\lambda_j=\lambda_{\rm obs}$, and $\lambda_j$ is the wavelength of transition $j$ in the rest frame. To implement this, we utilized Sherwood simulations \citep{bolton_sherwood_2017} to determine the effective optical depth for each Lyman series line across a range of absorption redshifts and $\Gamma$ values. We then include the proximity zone effect for each Lyman series line by matching the effective optical depth to a $\Gamma$ value that corresponds to $\Gamma_{\rm bg} + \Gamma_{\rm qso}$ as a function of distance from the QSO. 
We compute $\Gamma_{\rm qso}$ by dividing the line of sight into small steps of distance $\delta r$, and solving for $\Gamma_{\rm qso} (r)$ numerically using the method described in \citetalias{becker_mean_2021}. For the first step we assume that $\Gamma_{\rm qso}$ decreases purely geometrically, i.e.
\begin{equation}\label{eq:Gamma_qso_0}
    \Gamma_{\rm qso}(r=\delta r) = \Gamma_{\rm bg} \left( \frac{r}{R_{\rm eq}} \right)^{-2} \, .
\end{equation}
We then solve for $\Gamma_{\rm qso}(r + \delta r)$ over subsequent steps as 
\begin{equation}\label{eq:Gamma_qso_r}
    \Gamma_{\rm qso}(r + \delta r) = \Gamma_{\rm qso}(r) \left( \frac{r + \delta r}{r} \right)^{-2} e^{-\kappa_{\rm LL}(r) \delta r} \, ,
\end{equation}
where $\kappa_{\rm LL}(r)$ is computed using Equation~(\ref{eq:kappa}).
Finally, $\teff^j(z_j)$ and $\tau_{\rm eff}^{\rm Lyman} (\lambda_{\rm obs})$ can be computed for each combination of ($\kappa_{\rm LL}^{\rm bg}$, $\xi$, $z_{\rm qso}$, $R_{\rm eq}$), when fitting to the composite spectra. We have also tested that it does not significantly change the measured \mfp\ by either stacking the foreground Lyman series transmission based on $z_{\rm qso}$ and $R_{\rm eq}$ of each individual QSO (as in \citetalias{becker_mean_2021}), or computing a total foreground Lyman series transmission based on the averaged $z_{\rm qso}$ and $R_{\rm eq}$ in each redshift bin.

We use the same procedures for parameterizing the LyC transmission as outlined in \citetalias[][]{becker_mean_2021}. However, we make one modification by employing the recent measurements of $\Gamma_{\rm bg}$ from \citet{gaikwad_measuring_2023} that match multiple diagnostics of the IGM from observations to the \lya\ forest. For reference, the new estimates are $\Gamma_{\rm bg} \simeq 5\times10^{-13}\rm s^{-1}$ and $1.5\times10^{-13}\rm s^{-1}$ at $z=5.1$ and 6.0, respectively, in contrast to $7\times10^{-13}\rm s^{-1}$ and $3\times10^{-13}\rm s^{-1}$ utilized in \citetalias[][]{becker_mean_2021}. The new $\Gamma_{\rm bg}$ at $z=5.1$ is also consistent with measurements in e.g., \citet{daloisio_large_2018}. Moreover, instead of assuming a nominal $\pm 0.15$ dex error in $\Gamma_{\rm bg}$, we propagate the uncertainties in the measurements of $\Gamma_{\rm bg}$ from \citet{gaikwad_measuring_2023} into $R_{\rm eq}$. We discuss the effect of $\Gamma_{\rm bg}$ on \mfp\ in Section \ref{sec:dependence}.

We fit the transmission for each composite shown in Figure \ref{fig:composites} over 820--912~\AA\ in the rest frame.  Following \citetalias{becker_mean_2021}, uncertainties in \mfp\ are estimated using a bootstrap approach wherein we randomly draw QSO spectra with replacement in each redshift bin, and refit the new QSO composites for 10,000 realizations.  
To account for errors in redshift, we randomly shift the spectrum of each QSO that does not have a sub-mm $z_{\rm qso}$ in redshift following a Gaussian distribution with a standard deviation of 370 \kms (see Section \ref{sec:data}). We include the zero-point as a free parameter while fitting models to the composite, to account for flux zero-point errors.  We also treat the normalization of the LyC profile as a free parameter. We randomly vary $\xi$ by assuming a flat prior over $[0.3, 1.0]$. The fits are shown in the right-hand panel of Figure \ref{fig:composites}.

\section{Results and Discussion}\label{sec:results}

\subsection{\texorpdfstring{\mfp}{Lg} over \texorpdfstring{$5<z<6$}{Lg}}

\begin{figure}
\includegraphics[width=0.49\textwidth]{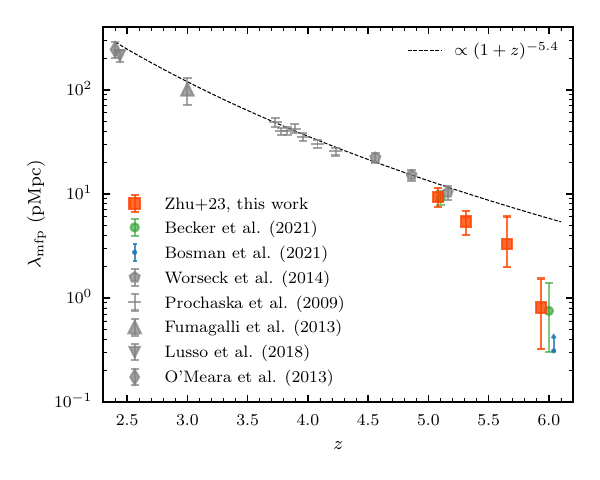}
\caption{
Direct measurements of \mfp\ from this work (orange-red squares) and from the literature \citep{becker_mean_2021,worseck_giant_2014,prochaska_direct_2009,fumagalli_dissecting_2013,lusso_spectral_2018,omeara_hstacswfc3_2013}. Error bars show 68\% limits. The dashed line shows the power-law extrapolation of \mfp\ from measurements at $z<5.16$ \citep{worseck_giant_2014}. The blue arrow shows the lower-limit constraint on the \mfp\ at $z=6$ from \citet{bosman_constraints_2021-1}. Symbols are displaced along redshift for display.
}
\label{fig:results}
\end{figure}

\begin{figure*}
\centering
\gridline{\fig{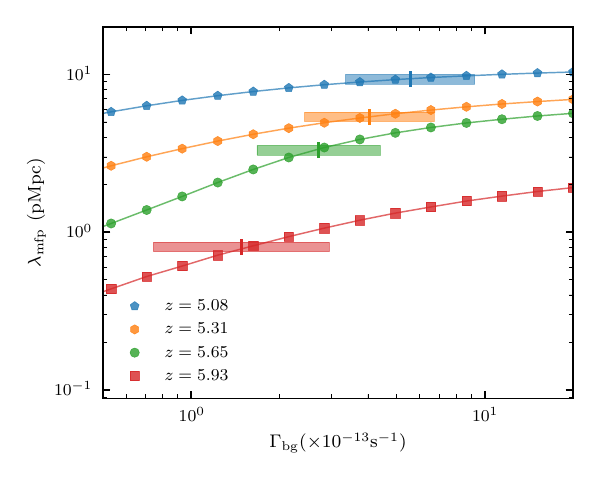}{0.49\textwidth}{(a)}
          \fig{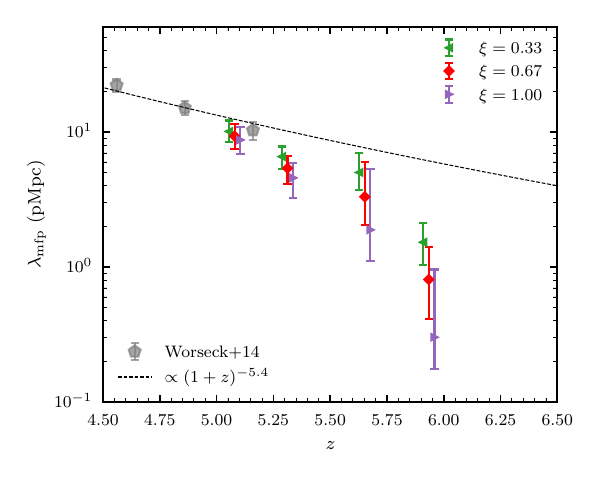}{0.49\textwidth}{(b)}}
\caption{
\textbf{(a)} Dependence of the \mfp\ measurements on $\Gamma_{\rm bg}$ at different redshifts. Nominal values and uncertainties of $\Gamma_{\rm bg}$ we adopt from \citet{gaikwad_measuring_2023} are shown with horizontal error bars.
\textbf{(b)} Measured \mfp\ based on fixed $\xi$ values of 0.33, 0.67, and 1.00. Symbols are displaced along redshift for display.
}
\label{fig:dependence}
\end{figure*}

We measure $\lambda_{\rm mfp} = 9.33_{-1.80(2.82)}^{+2.06(3.76)}$, $5.40_{-1.40(2.39)}^{+1.47(2.39)}$, $3.31_{-1.34(2.06)}^{+2.74(4.02)}$, and $0.81_{-0.48(0.68)}^{+0.73(1.22)}$ pMpc at the averaged redshifts $\langle z \rangle=5.08$, 5.31, 5.65, and 5.93, respectively.  The errors give 68\% (95\%) confidence limits.  Figure \ref{fig:results} plots our results along with previous direct \mfp\ measurements from the literature \citep[][and \citetalias{becker_mean_2021}]{prochaska_direct_2009,fumagalli_dissecting_2013,omeara_hstacswfc3_2013,worseck_giant_2014,lusso_spectral_2018}. Our measurements at both redshift ends are highly consistent with those presented in \citetalias[][]{becker_mean_2021}.
Our findings clearly indicate a rapid evolution in \mfp\ at $5<z<6$, particularly at $z\geq5.3$.

We have confirmed that using different redshift binning does not significantly affect our results. As the composite spectrum at $\langle z \rangle=5.65$ includes relatively fewer QSOs, we tested the robustness of our fitting using mock spectra. We created 1000 sets of $N=9$ mock QSO spectra with similar redshifts and $R_{\rm eq}$ as our sample. The mock spectra were based on our modeling of the transmission at $\lambda < 912$ \AA, with the mean free path randomly spanning a wide parameter space. We performed 1000 fitting realizations to each of these mock spectra sets, and found that the 68\% limits at $z=5.65$ could recover the simulated confidence level quite well. In addition, it is worth noting that none of the objects in our sample are identified as young QSOs with extremely small proximity zones \citep[e.g.,][]{eilers_detecting_2020,satyavolu_new_2023-1}, given that the \mfp\ values are quite short near $z \sim 6$.

We note that our $\langle z \rangle = 5.65$ stack (Figure \ref{fig:composites}) includes a small amount of transmission near $\lambda_{\rm rest}\sim 880$ \AA, even though the flux has been significantly attenuated at $\lambda_{\rm rest} > 890$ \AA.
We have inspected each individual spectrum in this redshift bin and have found potential transmission near this wavelength ($\lambda_{\rm obs}\sim 5800$ \AA) only in the ESI spectra, and not in the X-Shooter spectra. This region is at the boundary of two amplifiers of the ESI CCD.  It may also be contaminated by scattered light. This feature might therefore be an instrumental issue \footnote{\citet{prochaska_esikeck_2003} also report high flux near $\lambda_{\rm obs}\sim 5800$~\AA\ and infer that this is due to an incorrect matching in the gain of the two amplifiers for ESI. We still observe this feature after attempting to carefully account for the gain ratio, however.}; however, it is also possible that the transmission is real, in which case it may indicate a significant variation in the ionizing free paths towards individual spectra.  We hope to explore this in a future work.  For the uniform \mfp\ model used here, however, our tests based on  mock spectra indicate that this transmission feature does not significantly impact our measurements.

\subsection{Error analysis \& dependence on \texorpdfstring{$\Gamma_{\rm bg}$}{Lg} and \texorpdfstring{$\xi$}{Lg}} \label{sec:dependence}

\begin{table*}
\tablenum{2}
\caption{Error analysis for the measured \mfp}
\tabletypesize{\fontsize{10}{11}\selectfont}
\hspace{-0.5in}
\begin{tabular}{lrrrr}
\tableline \tableline
        \hspace{1.59cm} ~ \hspace{1.59cm}               & \hspace{1.59cm} $\langle z \rangle = 5.08$ \hspace{1.59cm} & \hspace{1.59cm} $\langle z \rangle = 5.31$ \hspace{1.59cm} & \hspace{1.59cm} $\langle z \rangle = 5.65$ \hspace{1.59cm} & \hspace{1.59cm}  $ \langle z \rangle =5.93$ \\
\hline
Measured $\lambda_{\rm mfp}$ & $9.33_{-1.80}^{+2.06}$     & $5.40_{-1.40}^{+1.47}$     & $3.31_{-1.34}^{+2.74}$     & $0.81_{-0.48}^{+0.73}$    \\
Fixed  $\Gamma_{\rm bg}$ and $\xi$ & $9.33_{-1.83}^{+1.95}$ & $5.40_{-1.14}^{+1.23}$ & $3.31_{-0.89}^{+2.52}$ & $0.81_{-0.34}^{+0.45}$    \\
Varying $\Gamma_{\rm bg}$ only & $9.33_{-0.69}^{+0.43}$ & $5.40_{-0.72}^{+0.62}$ & $3.31_{-0.99}^{+0.92}$ & $0.81_{-0.26}^{+0.21}$     \\
$\xi=0.33$                          & $10.10_{-1.69}^{+2.07}$ & $6.57_{-1.25}^{+1.23}$ & $5.02_{-1.32}^{+1.98}$ & $1.53_{-0.49}^{+0.59}$      \\
$\xi=0.67$                          & $9.33_{-1.81}^{+2.10}$ & $5.40_{-1.28}^{+1.27}$ & $3.31_{-1.26}^{+2.67}$ & $0.81_{-0.40}^{+0.60}$        \\
$\xi=1.00$                          &$8.74_{-1.88}^{+2.18}$ & $4.58_{-1.33}^{+1.30}$ & $1.89_{-0.77}^{+3.45}$ & $0.30_{-0.13}^{+0.66}$ \\ \tableline \tableline
\end{tabular}
\tablecomments{\mfp\ values are reported in pMpc.\\
(1) ``Measured \mfp'': our \mfp\ measurements at each redshift, including all sources of error;\\
(2) ``Fixed  $\Gamma_{\rm bg}$ and $\xi$'': statistical error from bootstrapping the QSO lines of sight without changing the nominal $\Gamma_{\rm bg}$ or $\xi$;\\
(3) ``Varying $\Gamma_{\rm bg}$ only'': \mfp\ values from varying $\Gamma_{\rm bg}$ while keeping the QSO lines of sight and $\xi$ fixed; \\
(4) Others: \mfp\ values based on different fixed $\xi$ values.
}
\label{tab:mfp_error}
\end{table*}

As described in Section \ref{sec:methods}, we mitigate the bias on \mfp\ from the QSO proximity effect by modeling its impact on the ionizing opacity.  The effect is parameterized by a nominal proximity zone size $R_{\rm eq}$, which specifies the proper distance at which the hydrogen ionization rate due to the QSO would be equal to the background rate in the absence of any attenuation. Therefore, the measured \mfp\ has some dependence on $R_{\rm eq}$.
Notably, the uncertainties in $R_{\rm eq}$ listed in Table \ref{tab:QSOlist} primarily emanate from $\Gamma_{\rm bg}$, with the contribution of uncertainty from $\alpha^{\rm UV}$ and $\alpha^{\rm ion}$ being relatively small ($\lesssim 10\%$). 
The measurements also depend on $\xi$, as suggested by Equation \ref{eq:kappa}.

Here, we have examined various sources of error in our \mfp\ measurements, including statistical error, error arising from $\Gamma_{\rm bg}$, and error stemming from $\xi$. Specifically, we evaluate the following: (1) the statistical error obtained by bootstrapping the QSO lines of sight while keeping the nominal values of $\Gamma_{\rm bg}$ and $\xi$ fixed, (2) \mfp\ values derived by varying $\Gamma_{\rm bg}$ while keeping the QSO composite and $\xi$ constant, and (3) \mfp\ values obtained for different fixed $\xi$. Table \ref{tab:mfp_error} summarizes the results. We would like to emphasize that fixing any of the parameters to their nominal value can lead to a slightly altered distribution of the resulting \mfp, and consequently, the corresponding 68\% limits may not align precisely with those of the main results. In this case, our primary interest lies in understanding the magnitude of the error.

The ``Fixed $\Gamma_{\rm bg}$ and $\xi$'' row indicates that the magnitude of the statistical error is comparable to the total error across all redshifts. This suggests that the dominant factor contributing to the error in our measurements is statistical fluctuations, encompassing factors such as the limited number of QSO spectra, flux noise, uncertainties in redshifts, and uncertainties in $\alpha_\nu^{\rm UV}$ and $\alpha_{\nu}^{\rm ion}$, among others. 
As shown in the third row, the random fluctuation in $\Gamma_{\rm bg}$ alone have a minor impact on the overall error. 
Regarding $\xi$, the last three rows indicate that using a lower (upper) value of $\xi$ shifts the \mfp\ estimates towards higher (lower) face values. This effect is comparable to the statistical fluctuations and is more pronounced at higher redshifts due to the relatively large $R_{\rm eq}$ in comparison to smaller \mfp. Thus, uncertainties in $\xi$ also make a significant contribution to our \mfp\ measurements.

We also explore how our \mfp\ measurements depend on $\Gamma_{\rm bg}$ and $\xi$ \emph{systematically}, such that the results can be easily adjusted for future  constraints. Figure \ref{fig:dependence}(a) illustrates the dependence of our best-fitting \mfp\ measurements on a wide range of $\Gamma_{\rm bg}$ values at each redshift. The figure also displays the nominal $\Gamma_{\rm bg}$ values and their uncertainties.  The dependence is minimal at $z \le 5.3$, where $\lambda_{\rm mfp} \propto \Gamma_{\rm bg}^{\sim 0.2}$. At these redshifts, the proximity zone size is smaller or comparable to \mfp, and hence, the measurements are not highly sensitive to $\Gamma_{\rm bg}$. At $z \ge 5.6$, however, $R_{\rm eq}$ is similar to or larger than \mfp, leading to a stronger dependence, with $\lambda_{\rm mfp}\propto \Gamma_{\rm bg}^{\sim 0.6}$. Nevertheless, if we adopt the higher end of $\Gamma_{\rm bg}=3\times10^{-13}\rm s^{-1}$ at $z=5.93$, \mfp\ would only increase to $\sim 1.0$ pMpc. The results would remain consistent with a steady and rapid \mfp\ evolution with time. 

While there might be an enhanced ionizing background due to  galaxies clustering near QSOs, recent research suggests this effect is likely secondary to QSO ionization. \citet{davies_ionization_2020} found that even the ``ghost'' proximity effect of QSOs --- a large-scale bias in the ionizing photon mean free path caused by QSO radiation --- could overwhelm the ionizing contribution from the clustering of nearby galaxies. In recent JWST observations, \citet{kashino_eiger_2023} also found that the impact of a QSO's ionizing radiation often dominates over local galactic sources near the QSO. These studies reinforce that, despite potential $\Gamma_{\rm bg}$ enhancements from galaxy clustering, the QSO's influence is typically predominant, as adopted in our modeling for the \mfp\ measurements.

For our main results, following \citetalias{becker_mean_2021}, the mean free path is measured based on a uniform distribution of $\xi \in [0.33, 1.00]$, and the face value is measured for $\xi=0.67$. As discussed in \citetalias{becker_mean_2021}, however, the scaling of $\kappa_{\rm LL}$ with $\Gamma$ is highly uncertain, especially for high redshifts. The scaling can be milder (smaller $\xi$) when neutral islands and/or self-shielding absorbers are present, and steeper (greater $\xi$) when the IGM shows a more uniform photoionization equilibrium \citep[see e.g.,][]{daloisio_hydrodynamic_2020,furlanetto_taxing_2005,mcquinn_lyman-limit_2011}. For reference, Figure \ref{fig:dependence}(b) shows how our measurements vary with fixed $\xi$ values of 0.33, 0.67, and 1.00. The face value and errors are also listed in Table \ref{tab:mfp_error}. Similar to the dependence on $\Gamma_{\rm bg}$, the measured \mfp\ becomes more sensitive to $\xi$ as redshift increases, and as the QSO's proximity effect becomes stronger relative to a smaller \mfp. Even with the extreme $\xi$ values discussed in \citetalias{becker_mean_2021}, nevertheless, the measurements are still consistent with our main results with the $1\sigma$ error bars, given the current data. 
Reassuringly, radiative transfer simulations recently developed by Roth et al. (in prep) suggest that the inferred \mfp\ using our methods only modestly depends ($\lesssim 20-30\%$) on the QSO lifetimes and environments (see also Satyavolu et al. in prep).
Future improved realistic reionization models would provide more insights into the scaling relation, especially when reionization is not fully concluded by $z=6$.

\subsection{Implication on when reionization ends}

Our measurements not only confirm the \mfp\ values presented in \citetalias[][]{becker_mean_2021} at $z=5.08$ and 5.93, but also depict a clear evolutionary trend over $5<z<6$. 
The mean free path increases steadily and rapidly with time: \mfp\ increases by a factor of $\sim4$ from $z\simeq6.0$ to $z\simeq5.6$, and by a factor of $\sim2$ every $\Delta z=0.3$ from $z\simeq5.6$ to $z\simeq 5.0$. This evolution carries important implications for the end of reionization. 

\citet{daloisio_hydrodynamic_2020} used radiative transfer hydrodynamic simulations to study the expected evolution of the mean free path following reionization.  They found that if reionization had ended well before $z=6$ and the IGM had sufficient time to relax hydrodynamically, then the evolution would expected to follow a trend of $\lambda_{\rm mfp} \propto (1+z)^{-5.4}$.  This relation, based on a fully ionized IGM with a homogeneous ionizing UVB, is identical to the best-fitting redshift dependence for direct \mfp\ measurements at $z \leq 5$ \citep{worseck_giant_2014}. However, as shown in Figure \ref{fig:results}, this relation significantly overpredicts the measurements by a factor of $\sim2-10$ over $5.3<z<6.0$. By this comparison, the data disfavor a fully ionized and relaxed IGM with a homogeneous UVB at these redshifts.

One possible explanation for the rapid evolution in \mfp\ is that reionization ends later than $z=6$.
Such a late-ending reionization scenario has recently been proposed to explain large-scale fluctuations in the observed \lya\ effective optical depth ($\teff$) at $z>5$ \citep[e.g.,][]{kulkarni_large_2019,keating_long_2020,nasir_observing_2020,choudhury_studying_2021,qin_reionization_2021}. The rapid evolution in \mfp\ can be naturally explained by ongoing reionization when large ionized bubbles merge and dense, optically-thick ionization sinks are photo-evaporated \citep[e.g.,][]{furlanetto_taxing_2005,sobacchi_inhomogeneous_2014,daloisio_hydrodynamic_2020}. As mentioned above, the rapid evolution in \mfp\ persists as late as $z\lesssim 5.3$, and the significant discrepancy between measurements and predictions from the relaxed IGM model also appears as late as $z=5.3$. Interestingly, the rapid evolution in \mfp\ appears to coincide in redshift with the appearance of large fluctuations in the observed \lya\ $\teff$ at $z \gtrsim 5.3$ \citep[e.g.,][]{becker_evidence_2015,eilers_opacity_2018,bosman_new_2018,bosman_hydrogen_2022,yang_measurements_2020}.
This may be due to the fact that a shorter \mfp, 
along with any potential neutral component from incomplete reionization, %
will boost the fluctuations in the ionizing UV background, producing scatter in $\teff$ \citep[e.g.,][]{davies_large_2016,nasir_observing_2020,qin_reionization_2021}. 
This joint evolution in the mean free path and UV background is expected near the end of reionization \citep[e.g.,][]{kulkarni_large_2019,keating_long_2020,nasir_observing_2020}.
Such a scenario is also consistent with long dark gaps observed in the \lya/\lyb\ forest \citep{zhu_chasing_2021,zhu_long_2022} at $z<6$, and the fraction of dark pixels measured in the forest \citep{mcgreer_model-independent_2015,jin_nearly_2023}.

\subsection{Comparison with reionization simulations}
\begin{figure*}
\centering
\gridline{\fig{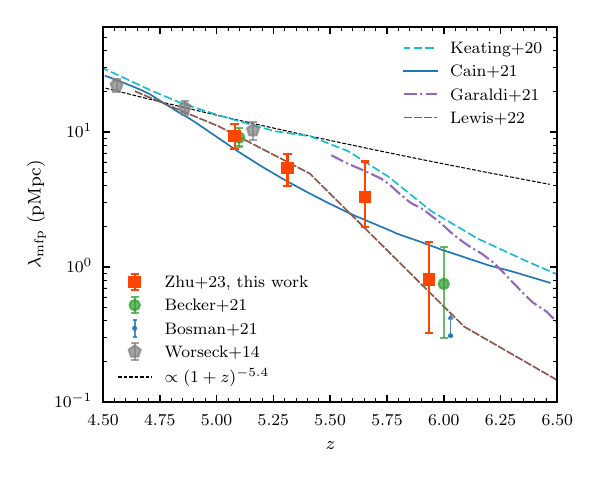}{0.49\textwidth}{(a)}
          \fig{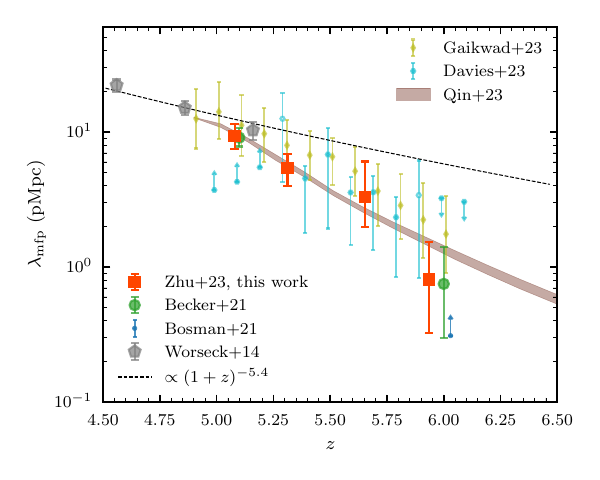}{0.49\textwidth}{(b)}}
\caption{
\textbf{(a)} Direct measurements of \mfp\ from this work (orange-red squares) compared to the predictions from recent models \citep{cain_short_2021,garaldi_thesan_2022,lewis_short_2022}. 
\textbf{(b)} Our measurements compared with indirect constraints based on \lya\ opacities \citep[][Davies et al.~in prep., Qin et al.~in prep.]{gaikwad_measuring_2023}. Open circles correspond to marginal constraints and arrows correspond to $2\sigma$ limits for Davies et al.~(in prep). The shaded region shows the posterior from Qin et al.~(in prep.) at the 68\% confidence level. In both panels, for comparison, we also show the direct \mfp\ measurements in \citet{becker_mean_2021} and \citet{worseck_giant_2014}, as well as the lower-limit constraint in \citet{bosman_constraints_2021-1}. Symbols are displaced along redshift for display.
}
\label{fig:discussion}
\end{figure*}

Figure \ref{fig:discussion}(a) compares our \mfp\ measurements to recent numerical simulations of reionization, including the enhanced-sink simulation in \citet[][]{cain_short_2021}, THESAN in \citet{garaldi_thesan_2022}, and CoDaIII in \citet[][]{lewis_short_2022}. These models use late-ending reionization histories and aim to explain the observed short \mfp\ at $z=6$ and the rapid evolution measured in \citetalias[][]{becker_mean_2021}, which we have confirmed in finer detail here.  We also include the ATON simulation (``low $\tau_{\rm CMB}$'' model) in \citet{keating_constraining_2020} for reference.

\citet[][]{cain_short_2021} reproduce \mfp\ that is consistent with the $z = 6$ measurements in a late-ending reionization driven by faint galaxies.  However, they also find that either a rapid drop in emissivity at $z < 6$ or extra sinks are required to reproduce the \mfp\ measurements at $z < 5.2$ in this scenario. %
\citet{garaldi_thesan_2022} use a radiative hydrodynamics simulation and generally reproduce the rapid evolution of \mfp\, although overshoot the $z=6$ measurement. They find that all of their late-reionization simulations can reproduce a dramatic evolution in \mfp\ from $z=5.5$ to 6, while one simulation wherein reionization ends by $z>6$ cannot. \citet{lewis_short_2022} also find that reionization ending later than $z<6$ is able to naturally explain the observations, although a drop in the emissivity is required near the end of reionization \citep{ocvirk_lyman-alpha_2021}. %

We note that these simulations measure the \mfp\ in slightly different manners. \citet[][]{cain_short_2021} generate mock LyC QSO spectra using randomly placed sightlines, and fit the stacked spectra using the model proposed by \citet{prochaska_direct_2009}. This procedure mimics the method used in \citet{worseck_giant_2014}. On the other hand, \citet{garaldi_thesan_2022} measure the distance at which the LyC transmission is attenuated by a factor of $1/e$ from individual sightlines, and take the average. Finally, \citet{lewis_short_2022} adopted multiple measurement methods for the \mfp, and the curve shown in Figure \ref{fig:discussion}(a) represents the median distance to a $1/e$ attenuation in LyC among sightlines. 
We also note that the models shown here may not necessarily reproduce observations of the \lya\ forest transmission \citep[e.g.,][]{garaldi_thesan_2022}. 
Nevertheless, the rapid evolution of \mfp\ that we measure over  $5 < z < 6$ is broadly consistent with these models that align with a late conclusion to reionization.

\subsection{Comparison with constraints on \texorpdfstring{\mfp}{Lg} from Ly\texorpdfstring{$\alpha$}{Lg} opacities}

Recently, \citet{gaikwad_measuring_2023} and Davies et al.~(in prep) have used alternative methods to probe the evolution of \mfp\ at these redshifts.  Instead of directly measuring \mfp\ from the LyC transmission, they use inference methods to jointly constrain \mfp\ and $\Gamma_{\rm bg}$ by modeling the observed \lya\ effective optical depth distribution.
Specifically, \citet{gaikwad_measuring_2023} post-processed hydrodynamic IGM simulations using a fluctuating UV background specified by a spatially-averaged  photoionization rate and a mean-free path parameter, $\lambda_0$.  These variables are constrained by comparing the simulated cumulative distribution function of $\teff$ with observations using a non-parametric two-sample Anderson-Darlington test.  A value of the physical mean free path, \mfp, is then inferred from the simulated neutral hydrogen distribution once the best-fitting UV background is applied.  This step is particularly significant at $z \lesssim 5.3$, where the \taueff\ distribution is consistent with a uniform UVB 
\citep[see also][]{becker_evidence_2018,bosman_hydrogen_2022} 
and does not constrain \mfp\ directly.  The fact that the constraints at these redshifts from \citet{gaikwad_measuring_2023} agree with the direct constrains presented here suggests that their 
simulations
may be modeling much of the ionizing opacity.

Davies et al.~(in prep) take a similar approach but use a combination of hydrodynamical and semi-numerical  simulations to model the density field, and employ a likelihood-free inference technique of approximate Bayesian computation to constrain \mfp\ and $\Gamma_{\rm bg}$ based on the \lya\ forest observations.  Davies et al.~(in prep) also 
constrain \mfp\ by treating it as an ``input'' sub-grid parameter for the UVB fluctuations rather than inferring it from a derived \ion{H}{1} density distribution.  This accounts for the fact that the Davies et al.~(in prep) values at $z < 5.3$ are lower limits.  %

As shown in Figure \ref{fig:discussion}(b), these indirect \mfp\ constraints are generally consistent ($\lesssim 1\sigma$) with the direct measurements presented here and in \citetalias{becker_mean_2021}.  This suggests that the rapidly evolving \mfp\ values needed for UV background fluctuations to drive the observed \lya\ \taueff\ distribution are consistent with the attenuation of ionizing photons we observe directly.  We note that \citet{gaikwad_measuring_2023} and Davies et al.~(in prep)  present somewhat different pictures of the IGM at these redshifts; the \citet{gaikwad_measuring_2023} models include neutral islands persisting to $z \simeq 5.2$ while Davies et al.~(in prep) have the flexibility to vary the \mfp\ distribution within ionized regions although no neutral islands are explicitly included. %
These models are broadly consistent with one another in that the \lya\ opacity fluctuations are mainly driven by UV background fluctuations, but this difference highlights the range of physical scenarios that are still formally consistent with the data.

We also include the inference based on multiple observations. Recently, Qin et al.~(in prep) use the {\tt 21cmFAST} Epoch of Reionization (EoR) simulations \citep[][]{mesinger_21cmfast_2011,murray_21cmfast_2020} to constrain IGM properties including its \mfp. Their input parameters represent galaxy properties such as the stellar-to-halo mass ratio, UV escape fraction and duty cycles. These allow them to evaluate the UV ionizing photon budget and forward model the impact on the IGM. Within a Bayesian framework, they include not only the XQR-30 measurement of the forest fluctuations \citep{bosman_hydrogen_2022} when computing the likelihood but also the CMB optical depth and galaxy UV luminosity functions. Therefore, the posterior they obtain for the IGM mean free path potentially %
represents a comprehensive constraint from all currently existing EoR observables. 
As Figure \ref{fig:discussion}(b) displays, the posterior from Qin et al.~(in prep) also show a rapid increase in the \mfp\ with time between $5\lesssim z \lesssim 6$. Although the posterior does not follow the exact trace of our direct measurements, the general consistency may suggest that such a rapid evolution in \mfp\ is potentially  favored by other EoR observables. 

\section{Conclusion \label{sec:conslusion}} 

We have presented new measurements of the ionizing mean free path between $z = 5.0$ and $6.0$.  These are the first direct measurements in multiple redshift bins over this interval, allowing us to trace the evolution of \mfp\ near the end of reionization.  Our measurements are based on new and archival data,
including new Keck/ESI observations and spectra from the XQR-30 program. By fitting the LyC transmission in composite spectra, we report \mfp = $9.33_{-1.80}^{+2.06}$, $5.40_{-1.40}^{+1.47}$, $3.31_{-1.34}^{+2.74}$, and $0.81_{-0.48}^{+0.73}$ pMpc, at $\langle z \rangle=5.08$, 5.31, 5.65, and 5.93, respectively. 
The results confirm the dramatic evolution in \mfp\ over $5<z<6$, as first reported in \citetalias[][]{becker_mean_2021}, and further show a steady and rapid evolution, with a factor of 
$\sim4$ increase from $z\simeq6.0$ to $z\simeq5.6$, and a factor of $\sim2$ increase every $\Delta z=0.3$ from $z\simeq5.6$ to $z\simeq 5.0$.
Our \mfp\ measurements disfavor a fully ionized and relaxed IGM with a homogeneous UVB at $\gtrsim 95\%$ confidence level down to at least $z\sim 5.3$ and are coincident with the onset of the fluctuations in observed $\teff$ at $z\sim 5.3$.

Recent indirect \mfp\ constraints based on IGM \lya\ opacity from \citet{gaikwad_measuring_2023} and Davies et al.~(in prep) agree well with our measurements and those in \citetalias{becker_mean_2021}. 
Our results are also broadly consistent with a range of late-ending reionization models \citep{cain_short_2021,garaldi_thesan_2022,lewis_short_2022,gaikwad_measuring_2023}.  Along with other probes from the \lya\ and \lyb\ forests, our results suggest that islands of neutral gas and/or large fluctuations in the UV background may persist in the IGM well below redshift six.

\begin{acknowledgments}

    We thank Joshua Roth and Ming-Feng Ho for helpful discussion. We thank the anonymous reviewer for their insightful comments. Y.Z., G.D.B., and H.M.C. were supported by the National Science Foundation through grant AST-1751404. H.M.C. was also supported by the National Science Foundation Graduate Research Fellowship Program under grant No. DGE-1326120. 
    A.D. acknowledges support from NASA 19-ATP19-0191, NSF AST-2045600, and JWST-AR-02608.001-A.
    S.E.I.B. acknowledges funding from the European Research Council (ERC) under the European Union’s Horizon 2020 research and innovation programme (grant agreement no. 740246 ``Cosmic Gas''). 
    G.K. is partly supported by the Department of Atomic Energy (Government of India) research project with Project Identification Number RTI~4002, and by the Max Planck Society through a Max Planck Partner Group.
    M.G.H. acknowledges the support of the UK Science and Technology Facilities Council (STFC) and the National Science Foundation under Grant No. NSF PHY-1748958. 
    Y.Q. acknowledges support from the Australian Research Council Centre of Excellence for All Sky Astrophysics in 3 Dimensions (ASTRO 3D), through project \#CE170100013.
    H.U. acknowledges support from JSPS KAKENHI grant No. 20H01953 and 22KK0231.

    Based on observations collected at the European Southern Observatory under ESO programme 1103.A-0817.

    Some of the data presented herein were obtained at the W. M. Keck Observatory, which is operated as a scientific partnership among the California Institute of Technology, the University of California and the National Aeronautics and Space Administration. The Observatory was made possible by the generous financial support of the W. M. Keck Foundation.
    The authors wish to recognize and acknowledge the very significant cultural role and reverence that the summit of Mauna Kea has always had within the indigenous Hawaiian community. We are most fortunate to have the opportunity to conduct observations from this mountain. Finally, this research has made use of the Keck Observatory Archive (KOA), which is operated by the W.M. Keck Observatory and the NASA Exoplanet Science Institute (NExScI), under contract with the National Aeronautics and Space Administration.

    This paper makes use of the following ALMA data: ADS/JAO.ALMA\allowbreak\#2022.1.00662.S, ADS/JAO.ALMA\allowbreak\#2021.1.01018.S, and ADS/JAO.ALMA\allowbreak\#2019.1.00111.S. ALMA is a partnership of ESO (representing its member states), NSF (USA) and NINS (Japan), together with NRC (Canada), MOST and ASIAA (Taiwan), and KASI (Republic of Korea), in cooperation with the Republic of Chile. The Joint ALMA Observatory is operated by ESO, AUI/NRAO and NAOJ. The National Radio Astronomy Observatory is a facility of the National Science Foundation operated under cooperative agreement by Associated Universities, Inc.

    This work has been supported by the Japan Society for the Promotion of Science (JSPS) Grants-in-Aid for Scientific Research (21H01128). This work has also been supported in part by the Sumitomo Foundation Fiscal 2018 Grant for Basic Science Research Projects (180923), the Collaboration Funding of the Institute of Statistical Mathematics ``New Perspective of the Cosmology Pioneered by the Fusion of Data Science and Physics''.

    For the purpose of open access, the authors have applied a Creative Commons Attribution (CC BY) licence to any Author Accepted Manuscript version arising from this submission.

\facilities{ALMA, Keck:II (ESI), VLT:Kueyen (X-Shooter)}

\software{
    {\tt Astropy} \citep{astropy_collaboration_astropy_2013},
    {\tt CASA} \citep{mcmullin_casa_2007,casa_team_casa_2022},
    {\tt GDL} \citep{coulais_scaling_2014}, 
    {\tt Julia \citep{bezanson_julia_2015}},
    {\tt Matplotlib} \citep{hunter_matplotlib_2007},
    {\tt NumPy} \citep{van_der_walt_numpy_2011},
    {\tt SpectRes} \citep{carnall_spectres_2017}
}
\end{acknowledgments}

\vspace{1in}

\end{document}